\documentclass[prd,aps,10pt,notitlepage,showpacs,nofootinbib,tightenlines]{revtex4}
\usepackage{mathrsfs}
\usepackage{amsmath}
\usepackage{amssymb}
\usepackage{epsfig}
\usepackage{graphicx}
\usepackage{booktabs}
\usepackage{multirow}
\usepackage{subfigure}
\usepackage{bm}
\usepackage{times}
\usepackage{braket}
\usepackage{color}
\usepackage{slashed}
\usepackage{hyperref}
\usepackage{threeparttable}
\newcommand{\beq}{\begin{eqnarray}}
\newcommand{\eeq}{\end{eqnarray}}
\newcommand{\non}{\nonumber\\ }

\definecolor{Red}{rgb}{1.,0.,0.}

\definecolor{Yellow}{rgb}{0.,1.,0.}

\definecolor{Blue}{rgb}{0.,0.,1.}

\definecolor{nicered}{rgb}{0.7,0.1,0.1}
\definecolor{nicegreen}{rgb}{0.1,0.5,0.1}
\hypersetup{colorlinks,citecolor=nicegreen,linkcolor=nicered}

\def \epjc{ Eur. Phys. J. C }

\def \npb{  Nucl. Phys. B }
\def \plb{  Phys. Lett. B }

\def \prd{  Phys. Rev. D }
\def \prl{  Phys. Rev. Lett.  }

\def \jhep{ J. High Energy Phys. }

\begin{document}

\title{Study of four-body decays $B_{(s)} \to (\pi\pi)(\pi\pi)$ in the perturbative QCD approach}

\author{Da-Cheng Yan$^1$}         
\author{Zhou Rui$^2$}              
\author{Yan Yan$^{1}$}    
\author{Ya Li$^3$}                \email[Corresponding author:]{liyakelly@163.com}
\affiliation{$^1$ School of Mathematics and Physics, Changzhou University, Changzhou, Jiangsu 213164, China}
\affiliation{$^2$ College of Sciences, North China University of Science and Technology, Tangshan, Hebei 063210, China}
\affiliation{$^3$ Department of Physics, College of Sciences, Nanjing Agricultural University, Nanjing, Jiangsu 210095, China}
\date{\today}

\begin{abstract}
In this work, we make a systematical study on the four-body $B_{(s)} \to (\pi\pi)(\pi\pi)$ decays in the perturbative QCD (PQCD) approach,
where the $\pi\pi$ invariant mass spectra are dominated by the vector resonance $\rho(770)$
and the scalar resonance $f_0(980)$.
We improve the Gengenbauer moments for the longitudinal $P$-wave two-pion distribution amplitudes (DAs)
by fitting the PQCD factorization formulas to measured branching ratios of three-body and four-body $B$ decays.
With the fitted Gegenbauer moments, we make predictions for the branching ratios and direct $CP$ asymmetries of four-body $B_{(s)}  \to (\pi\pi)(\pi\pi)$ decays.
As a by-product, we extract the branching ratios of two-body $B_{(s)}  \to \rho\rho$ from the corresponding four-body
decay modes and calculate the relevant polarization fractions.
We find that the ${\cal B}(B^0 \to \rho^+\rho^-)$ is consistent with the previous theoretical predictions and data.
The leading-order PQCD calculations of the ${\cal B}(B^+\to \rho^+\rho^0)$, ${\cal B}(B^0\to \rho^0\rho^0)$ and the $f_0(B^0\to \rho^0\rho^0)$ are a bit lower than the experimental measurements, which should be further examined.
It is shown that the direct $CP$ asymmetries are large when both tree and penguin contributions are comparable to each other,
but small for the tree-dominant or penguin-dominant processes.
In addition to the direct $CP$ asymmetries,
the ``true" and ``fake" triple-product asymmetries (TPAs) originating from the interference
among various helicity amplitudes in the $B_{(s)}\to (\pi\pi)(\pi\pi)$ decays are also analyzed.
The sizable averaged TPA ${\cal A}_{\text{T-true}}^{1, \text{ave}}=25.26\%$ of the color-suppressed decay $B^0\to \rho^0\rho^0 \to (\pi^+\pi^-)(\pi^+\pi^-)$ is predicted for the first time, which deviates a lot from the so-called ``true" TPA $\mathcal{A}_\text{T-true}^1=7.92\%$ due to the large direct $CP$ violation.
A large ``fake" TPA $\mathcal{A}_\text{T-fake}^1=24.96\%$ of the decay $B^0\to \rho^0\rho^0 \to (\pi^+\pi^-)(\pi^+\pi^-)$ is also found, which indicates the significance of the final-state interactions.
The predictions in this work can be tested by LHCb and Belle-II experiments in the near future.
\end{abstract}

\maketitle

\section{Introduction}
The study of the charmless non-leptonic $B$ meson decays can offer rich opportunities
to understand the underlying mechanism for hadronic weak decays and $CP$ violation,
to test the standard model (SM) by measuring the related Cabibbo-Kobayashi-Maskawa (CKM) matrix parameters such as the weak phases $\alpha,\beta$, and $\gamma$,
and also to search for the possible signals of new physics (NP) beyond the SM.
The precise measurement of the phase $\alpha$ of the CKM unitarity triangle is a hot topic in the particle physics,
and have drawn great attention for both theorists and experimentalists.
In principle, direct experimental measurements of angle $\alpha$ can be made from decays that proceed mainly through a
${\bar b\to u{\bar u} d}$ tree diagram, like $B^0\to \rho^+\rho^-$ decay.
However, the additional presence of the penguin contributions complicates the measurement of $\alpha$ from such decay.
At present, one of the most favorable methods to determine the CKM phase $\alpha$ is through an isospin analysis of the
$B^0 \to \rho^{+}\rho^{-}, \rho^{0}\rho^{0}$ and $B^+\to \rho^+\rho^0$ decays ~\cite{pdg2020}.
Therefore, a much deeper understanding of the related phenomena is imperative.

In $B\to \rho\rho$ decay channels,
a spin zero particle decays into two spin one particles,
and subsequently  each $\rho$ meson decays into a $\pi\pi$ pair.
The corresponding decay amplitude can be written as superposition of three helicity amplitudes,
including one longitudinal polarization $A_0$ and two transverse polarization with spins parallel $A_\|$ or perpendicular $A_\bot$ to each other.
The longitudinal amplitude $A_0$ and one of the transverse amplitude $A_\|$ are $CP$-even,
while another transverse amplitude $A_\bot$ is $CP$-odd.
Besides the direct $CP$ asymmetries,
the interferences between the $CP$-even and $CP$-odd amplitudes
can generate $CP$ asymmetries in angular distributions,
which is the so-called triple product asymmetries (TPAs)~\cite{prd-393339,npbps-13487,ijmpa-192505}.
A scalar triple product takes the form,
$\vec{v}_1\cdot (\vec{v}_2 \times \vec{v}_3)$,
with each $\vec{v}_i$ being a spin or momentum of the final-state particle.
Obviously, this triple product is odd under the time reversal ($T$) transformation,
and also under the $CP$ transformation due to the CPT invariance.
Experimentally,
TPAs have already been measured by Belle, $BABAR$ , CDF
and LHCb Collaborations~\cite{prl95-091601,prd76-031102,jhep07-166,prl107-261802,jhep05-026,plb713-369,prd90-052011,LHCb:2013xyz,LHCb:2014xzf,LHCb:2019jgw}.

As is well known, the $CP$ violation originates from the weak phase in the CKM quark-mixing matrix in the SM.
A non-zero direct $CP$ asymmetry generally requires the interference of at least two amplitudes with a weak phase difference $\Delta \phi$ and a strong phase
difference $\Delta \delta$, and is proportional to $\sin\Delta \phi \sin\Delta \delta $.
As claimed in Ref.~\cite{ijmpa-192505},
strong phases should always be rather small in the $B$ decays since the $b$-quark is heavy.
Thus, the direct $CP$ asymmetries are usually highly suppressed by the term $\sin\Delta \delta $.
The TPAs  going as $\sin\Delta \phi \cos\Delta \delta$, nevertheless, is maximal when $\Delta \delta $ is close to zero.
It indicates that direct $CP$ violation and TPAs can complement each other.
Particularly,
even when the effect of $CP$ violation is absent,
$T$-odd triple products (also called the ``fake" TPAs) proportional to $\cos\Delta \phi \sin \Delta \delta$
can provide further insight on new physics~\cite{plb701-357}.
Phenomenological investigations on these asymmetries  have been conducted in Refs.~\cite{prd-88016007,prd84-096013,plb701-357,prd-86076011,prd-92076013,prd-87116005,2211.07332}.

\begin{figure}[tbp]
	\centering
	\includegraphics[scale=0.7]{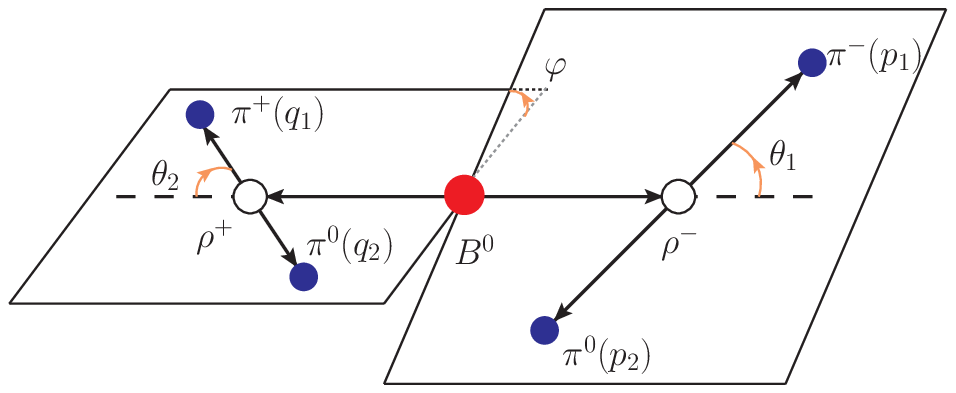}
\vspace{-14cm}
	\caption{Graphical definitions of the helicity angles $\theta_1$, $\theta_2$ and $\varphi$ for the $B^0 \to \rho^+\rho^-$ decay, with each quasi-two-body intermediate resonance decaying to two pseudoscalars ($\rho^+ \to \pi^+\pi^0$ and $\rho^- \to \pi^-\pi^0$).
	$\theta_{1}(\theta_{2})$ is denoted as the angle between the direction of motion of $\pi^{-}(\pi^{+})$ in the $\rho^{-}(\rho^{+})$ rest frame and $\rho^{-}(\rho^{+})$ in the $B^0$ rest frame, and $\varphi$ is the angle between the two planes defined by $\pi^+\pi^0$ and $\pi^-\pi^0$ in the $B^0$ rest frame.}
\label{fig1}
\end{figure}

\begin{figure}[tbp]
	\centering
	\includegraphics[scale=0.8]{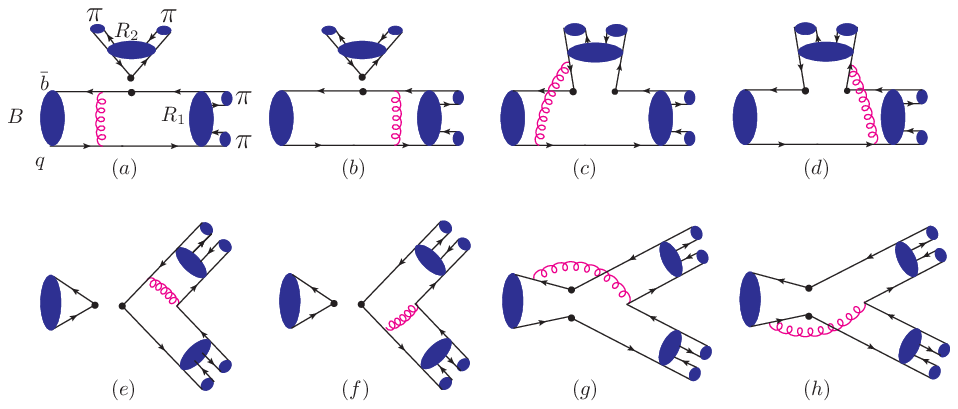}
\vspace{-16cm}
	\caption{Typical leading-order Feynman diagrams for the four-body decays $B \to R_1R_2\to (\pi\pi)(\pi\pi)$ with $q=(d,s)$, where the symbol $\bullet$ denotes a weak interaction vertex. The diagrams ($a$)-($d$) represent the $B\to (R_1\to) \pi\pi$ transition, as well as the diagrams ($e$)-($h$) for annihilation contributions.}
\label{fig2}
\end{figure}

The construction of the above mentioned TPAs usually requires at least four particles in the final states.
In the previous theoretical studies,
the $B\to \rho\rho$ decays are usually treated as two-body processes,
and have been analyzed in detail in the
two-body framework by employing the QCD factorization (QCDF)~\cite{npb774-64,prd80-114026,npb990-116175}, the perturbative QCD (PQCD) approaches~\cite{prd76-074018,prd91-054033,npb935-17},
the soft-collinear effective theory (SCET)~\cite{prd96-073004}, and the factorization-assisted topological amplitude approach (FAT)~\cite{epjc77-333}.
Whereas,
since the vector resonance $\rho$ is always detected via its decay $\rho \to \pi\pi$,
they are at least four-body decays on the experimental side as shown in Fig.~\ref{fig1}.
The related measurements have been reported by $BABAR$ and Belle Collaborations~\cite{prd-052007,prd-071104,prl-141802,prl-221801,prd-072008,prd-032010}.
Therefore,
we intend to analyze the four-body decays $B_{(s)} \to (\pi\pi)(\pi\pi)$ in the PQCD approach based on $k_T$ factorization in the present work
with the relevant Feynman diagrams illustrated in Fig.~\ref{fig2}.
The invariant mass of the $\pi\pi$ pair is restricted to be in the range of $0.3<\omega_{\pi\pi}<1.1$ GeV~\cite{jhep05-026}.
Except for the dominant vector resonance $\rho$, the important scalar background  $f_0(980)\to \pi\pi$ is also taken into account.
Throughout the remainder of the paper,
the symbol $f_0$ is used to denote the scalar $f_0(980)$ resonance.
We work out a number of physical observables, including branching ratios, polarization fractions as well as a set of $CP$ asymmetries.
Specially, the TPAs of the $B_{(s)} \to (\pi\pi)(\pi\pi)$ decays have been predicted for the first time  in the PQCD approach.

To be honest, the strong dynamics of the multi-body  $B$ meson decays is indeed much more complicated than two-body ones,
since both resonant and nonresonant contributions can contribute together with substantial final-state interactions (FSIs)~\cite{prd89-094013,1512-09284,prd89-053015}.
Therefore, the factorization formalism that describes a multi-body decay in full phase space has not been established so far.
As a first step,
we can only restrict ourselves to the specific kinematical configurations in which each two particles fly collinearly and two pairs of final
state particles recoil back in the rest frame of the $B$ meson, see Fig.~\ref{fig1}.
The production of the final state meson pairs can then be factorized into the non-perturbative two-hadron distribution amplitudes (DAs)~\cite{MP,MT01,MT02,MT03,NPB555-231,Grozin01,Grozin02}.
Thereby, the total decay amplitude $\cal A$ of the corresponding $B_{(s)}\to (\pi\pi)(\pi\pi)$ four-body decays in PQCD can be described in terms of
\begin{eqnarray}
\mathcal{A}=\Phi_B\otimes H\otimes \Phi_{\pi\pi}\otimes\Phi_{\pi\pi},
\label{deam}
\end{eqnarray}
where $\Phi_B$ is the $B$ meson DA, and the two-hadron DA $\Phi_{\pi\pi}$ absorbs the nonperturbative dynamics involved in the $\pi\pi$ meson pair.
The hard kernel $H$ describes the dynamics of the strong and electroweak interactions in four-body hadronic decays in a similar way as the one for the corresponding two-body decays.
The symbol $\otimes$ denotes the convolution of the above factors in parton momenta.
In our calculations,
we adopt the widely used wave function for $B$ meson in the PQCD factorization approach~\cite{Li:2003yj,Kurimoto:2001zj}.
For more alternative models of the $B$ meson DA and the subleading contributions,
one can refer to Refs.~\cite{prd102011502,prd70074030,Li:2012md,Li:2014xda,Li:2012nk,2207.02691,jhep05-157}.
The $S$- and $P$-wave contributions are parametrized into the corresponding time-like form factors involved in the two-meson DAs.
The longitudinal polarization $P$-wave $\pi\pi$ DAs $\Phi_{\pi\pi}$ have been extracted from a
global analysis of three-body charmless hadronic decays $B_{(s)}\to P\rho\to P\pi\pi$, with $P=\pi, K$,  in the PQCD approach~\cite{2105-03899}.
We will improve the fitting results with the inclusion of the currently available data for the four-body decays  $B_{(s)}\to \rho\rho\to (\pi\pi)(\pi\pi)$ in the following section.
On the theoretical side, an important breakthrough of four-body $B$ meson decays has been achieved based on the quasi-two-body-decay mechanism.
For example,
within the framework of the quasi-two-body QCDF approach,
the localized $CP$ violation and branching fraction of the four-body decay
$\bar{B}^0\to K^-\pi^+\pi^+\pi^-$ have been calculated in Refs.~\cite{1912-11874,2008-08458}.
Besides,
the PQCD factorization formalism relying on the quasi-two-body decay mechanism for four-body $B$ meson decays has also been well established in Refs.~\cite{zjhep,Li:2021qiw,prd105-053002,prd105-093001,Liang:2022mrz}.

The rest of the paper is organized as follows. The kinematic variables for four-body hadronic
$B$ meson decays are defined in Sec.\ref{sec2}.
The considered two-meson $P$-wave and $S$-wave DAs are also
parametrized, whose normalization form factors are assumed to adopt the Gounaris-Sakurai (GS)
model~\cite{prl21-244} and Flatt\'e model~\cite{plb63-228}, respectively.
We explain how to perform the global fit, present and discuss the numerical results in Sec.\ref{sec3}, which is followed by the Conclusion.
The Appendix collects the explicit PQCD factorization formulas for all the decay amplitudes.

\section{Framework}\label{sec2}
\subsection{Kinematics}
Taking the four-body decay $B^0\to \rho^+\rho^-\to (\pi^+\pi^0)(\pi^-\pi^0)$ displayed in Fig.~\ref{fig1} as an example,
the external momenta of the decay chain will be denoted as $p_B$ for the $B$ meson,
$p(q)$ for the intermediate $\rho^{-}(\rho^+)$ resonance,
and $p_{i} (i=1-4)$ for the four final-state $\pi$ mesons,
with $p_{B}=p+q$, $p=p_1+p_2$,  $q=p_3+p_4$  obeying the momentum conservation.
As usual, we shall consider the $B$ meson at rest and work in the light-cone coordinates,
such that
\begin{eqnarray}
p_{B}&=&\frac{m_{B}}{\sqrt 2}(1,1,\textbf{0}_{\rm T}),\quad\quad
p=\frac{m_{B}}{\sqrt2}(g^+,g^-,\textbf{0}_{\rm T}),\quad\quad
q=\frac{m_{B}}{\sqrt 2}(f^-,f^+,\textbf{0}_{\rm T}),
\end{eqnarray}
where $m_B$ represents the $B$ meson mass.
The factors $f^{\pm},g^{\pm}$ related to the invariant masses of the meson pairs via $p^2=\omega_1^2$ and $q^2=\omega_2^2$
can be written as
\begin{eqnarray}\label{eq:epsilon}
g^{\pm}&=&\frac{1}{2}\left[1+\eta_1-\eta_2\pm\sqrt{(1+\eta_1-\eta_2)^2-4\eta_1}\right],\nonumber\\
f^{\pm}&=&\frac{1}{2}\left[1-\eta_1+\eta_2\pm\sqrt{(1+\eta_1-\eta_2)^2-4\eta_1}\right],
\end{eqnarray}
with the mass ratios $\eta_{1,2}=\frac{\omega_{1,2}^2}{m^2_{B}}$.
For the evaluation of the hard kernels $H$ in the PQCD approach,
we also need to define three valence quark momenta labeled by $k_i$ $(i=B,p,q)$ in each meson as
\begin{eqnarray}
k_{B}&=&\left(0,x_B p_B^+ ,\textbf{k}_{B \rm T}\right),\quad
k_p= \left( x_1 p^+,0,\textbf{k}_{1{\rm T}}\right),\quad\quad
k_q=\left(0,x_2q^-,\textbf{k}_{2{\rm T}}\right),\label{eq:mom-B-k}
\end{eqnarray}
with the parton momentum fraction $x_i$, and the parton transverse momentum $\textbf{k}_{\rm iT}$.
We have dropped the small component $k^-_{p}$ ( $k^+_{q}$) in Eq.~\ref{eq:mom-B-k}
because $k_p (k_q )$ is aligned with the meson pair in the plus (minus) direction.
The plus component of the momenta $k_B$ is also neglected since it
does not appear in the hard kernels for dominant factorizable contributions.
The corresponding longitudinal polarization vectors of the $P$-wave $\pi\pi$ pairs are defined as
 \begin{eqnarray}\label{eq:pq1}
\epsilon_{p}=\frac{1}{\sqrt{2\eta_1}}(g^+,-g^-,\textbf{0}_{\rm T}),\quad
\epsilon_{q}=\frac{1}{\sqrt{2\eta_2}}(-f^-,f^+,\textbf{0}_{\rm T}),
\end{eqnarray}
which satisfy the normalization $\epsilon_{p}^2=\epsilon_{q}^2=-1$  and the orthogonality
$\epsilon_{p}\cdot p=\epsilon_{q}\cdot q=0$.

Based on the relations $p=p_1+p_2$ and $q=p_3+p_4$, as well as the on-shell conditions $p_i^2=m_i^2$ for the final state mesons,
one can derive the explicit expressions of the individual momentum $p_i$
\begin{eqnarray}\label{eq:p1p4}
p_1&=&\left(\frac{m_{B}}{\sqrt{2}}(\zeta_1+\frac{r_1-r_2}{2\eta_1})g^+,\frac{m_{B}}{\sqrt{2}}(1-\zeta_1+\frac{r_1-r_2}{2\eta_1})g^-,\textbf{p}_{\rm T}\right),\nonumber\\
p_2&=&\left(\frac{m_{B}}{\sqrt{2}}(1-\zeta_1-\frac{r_1-r_2}{2\eta_1})g^+,\frac{m_{B}}{\sqrt{2}}(\zeta_1-\frac{r_1-r_2}{2\eta_1})g^-,-\textbf{p}_{\rm T}\right),\nonumber\\
p_3&=&\left(\frac{m_{B}}{\sqrt{2}}(1-\zeta_2+\frac{r_3-r_4}{2\eta_2})f^-,\frac{m_{B}}{\sqrt{2}}(\zeta_2+\frac{r_3-r_4}{2\eta_2})f^+,\textbf{q}_{\rm T}\right),\nonumber\\
p_4&=&\left(\frac{m_{B}}{\sqrt{2}}(\zeta_2-\frac{r_3-r_4}{2\eta_2})f^-,\frac{m_{B}}{\sqrt{2}}(1-\zeta_2-\frac{r_3-r_4}{2\eta_2})f^+,-\textbf{q}_{\rm T}\right),\nonumber\\
\textbf{p}_{\rm T}^2&=&\zeta_1(1-\zeta_1)\omega_1^2+\alpha_1,\quad
\textbf{q}_{\rm T}^2=\zeta_2(1-\zeta_2)\omega_2^2+\alpha_2,
\end{eqnarray}
with the factors
\begin{eqnarray}\label{eq:alpha12}
\alpha_1=-\frac{r_1+r_2}{2\eta_1}+\frac{(r_1-r_2)^2}{4\eta_1^2},\quad
\alpha_2=-\frac{r_3+r_4}{2\eta_2}+\frac{(r_3-r_4)^2}{4\eta_2^2},
\end{eqnarray}
and the ratios $r_i=m_i^2/m^2_B$, $m_i$ being the masses of the final state mesons.
The factors $\zeta_1+\frac{r_1-r_2}{2\eta_1}=p_1^+/p^+$ and $\zeta_2+\frac{r_3-r_4}{2\eta_2}=p_3^-/q^-$ characterize the momentum fraction for one of pion-pion pair.

The relation between $\zeta_{1,2}$ and the polar angle $\theta_{1,2}$ in the dimeson rest frame in Fig.~\ref{fig1} can be obtained easily,
\begin{eqnarray}\label{eq:cos}
2\zeta_{1}-1=\sqrt{1-2\frac{r_1+r_2}{\eta_1}+\frac{(r_1-r_2)^2}{\eta_1^2}}\cos\theta_1,\nonumber\\
2\zeta_{2}-1=\sqrt{1-2\frac{r_3+r_4}{\eta_2}+\frac{(r_3-r_4)^2}{\eta_2^2}}\cos\theta_2,
\end{eqnarray}
with the upper and lower limits of $\zeta_{1,2}$
\begin{eqnarray}
\zeta_{1\text{max,min}}=\frac{1}{2}\left[1\pm\sqrt{1+4\alpha_1}\right],~\quad
\zeta_{2\text{max,min}}=\frac{1}{2}\left[1\pm\sqrt{1+4\alpha_2}\right].
\end{eqnarray}

\subsection{Two-meson Distribution amplitudes}
In the quasi-two-body framework, the resonant contributions through two-body channels can be included by parameterizing the two-meson DAs along with the corresponding time-like form factors.
In this section, we will briefly introduce the $S$- and $P$-wave two-pion DAs, as well as the time-like form factors used in our work.
In what follows the subscripts $S$ and $P$ are always related to the corresponding partial waves.

The $S$-wave two-pion DA can be written in the following form~\cite{prd91-094024},
\begin{eqnarray}\label{swave}
\Phi_{S}(z,\omega)=\frac{1}{\sqrt{2N_c}}[{p\hspace{-1.5truemm}/}\phi^0_S(z,\omega^2)+
\omega\phi^s_S(z,\omega^2)+\omega({n\hspace{-2.0truemm}/}{v\hspace{-2.0truemm}/}-1)\phi^t_S(z,\omega^2)],
\end{eqnarray}
in which $N_c$ is the number of colors, and the asymptotic forms of the individual twist-2 and twist-3 components $\phi_S^0$ and $\phi_S^{s,t}$
are parameterized as~\cite{MP,MT01,MT02,MT03}
\begin{eqnarray}
\phi^0_S(z,\omega^2)&=&\frac{9F_S(\omega^2)}{\sqrt{2N_c}}a_S z(1-z)(1-2z),\label{eq:phis0}\\
\phi^s_S(z,\omega^2)&=&\frac{F_S(\omega^2)}{2\sqrt{2N_c}},\\
\phi^t_S(z,\omega^2)&=&\frac{F_S(\omega^2)}{2\sqrt{2N_c}}(1-2z),
\end{eqnarray}
with the time-like scalar form factor $F_S(\omega^2)$.
The Gegenbauer moment $a_S$ in Eq.~(\ref{eq:phis0}) is adopted the same value as that in Ref.~\cite{epjc76-675}: $a_S=0.2\pm 0.2$.

\begin{figure}[tbp]
	\centering
	\includegraphics[scale=0.27]{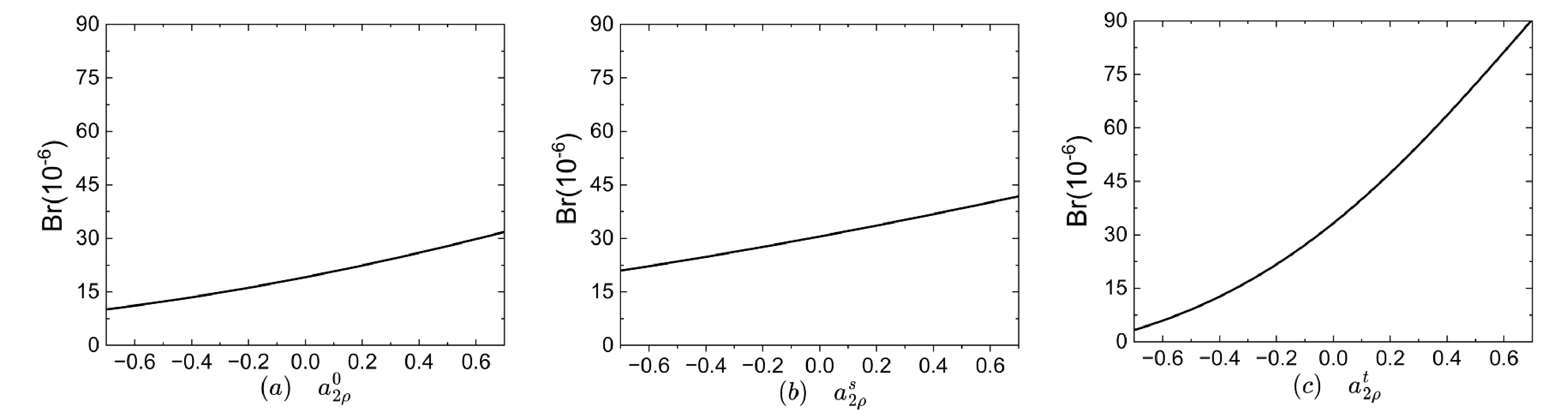}
	\caption{The longitudinal branching ratio of the four-body decay $B^0\to \rho^+\rho^-\to (\pi^+\pi^0)(\pi^-\pi^0)$ as a function of the Gegenbauer coefficients
: $(a)$ $ a^0_{2\rho}$,  $(b)$ $a^s_{2\rho}$, and  $(c)$ $a^t_{2\rho}$.}
\label{a0st}
\end{figure}
The corresponding $P$-wave two-pion DAs related to both longitudinal and transverse polarizations are decomposed,
up to the twist 3, into~\cite{prd98-113003}:
\begin{eqnarray}
\Phi_P^{L}(z,\zeta,\omega)&=&\frac{1}{\sqrt{2N_c}} \left [{ \omega \epsilon\hspace{-1.5truemm}/_p  }\phi_P^0(z,\omega^2)+\omega\phi_P^s(z,\omega^2)
+\frac{{p\hspace{-1.5truemm}/}_1{p\hspace{-1.5truemm}/}_2
  -{p\hspace{-1.5truemm}/}_2{p\hspace{-1.5truemm}/}_1}{\omega(2\zeta-1)}\phi_P^t(z,\omega^2) \right ] (2\zeta-1)\;,\label{pwavel}\\
\Phi_P^{T}(z,\zeta,\omega)&=&\frac{1}{\sqrt{2N_c}}
\Big [\gamma_5{\epsilon\hspace{-1.5truemm}/}_{T}{ p \hspace{-1.5truemm}/ } \phi_P^T(z,\omega^2)
+\omega \gamma_5{\epsilon\hspace{-1.5truemm}/}_{T} \phi_P^a(z,\omega^2)+ i\omega\frac{\epsilon^{\mu\nu\rho\sigma}\gamma_{\mu}
\epsilon_{T\nu}p_{\rho}n_{-\sigma}}{p\cdot n_-} \phi_P^v(z,\omega^2) \Big ]\non
&&\cdot \sqrt{\zeta(1-\zeta)+\alpha_1}\label{pwavet}\;.
\end{eqnarray}
The various twists $\phi_P^i$ in the above equations can be expanded in terms of the Gegenbauer polynomials:
\begin{eqnarray}
\phi_{P}^0(z,\omega^2)&=&\frac{3F_{P}^{\parallel}(\omega^2)}{\sqrt{2N_c}}z(1-z)\left[1
+a^0_{2\rho}\frac{3}{2}(5(1-2z)^2-1)\right] \;,\label{eq:phi0}\\
\phi_{P}^s(z,\omega^2)&=&\frac{3F_{P}^{\perp}(\omega^2)}{2\sqrt{2N_c}}(1-2z)\left[1
+a^s_{2\rho}(10z^2-10z+1)\right]  \;,\\
\phi_{P}^t(z,\omega^2)&=&\frac{3F_{P}^{\perp}(\omega^2)}{2\sqrt{2N_c}}(1-2z)^2\left[1
+a^t_{2\rho}\frac{3}{2}(5(1-2z)^2-1)\right]  \;,\label{eq:phit}\\
\phi_{P}^T(z,\omega^2)&=&\frac{3F_{P}^{\perp}(\omega^2)}
{\sqrt{2N_c}}z(1-z)[1+a^{T}_{2\rho}\frac{3}{2}(5(1-2z)^2-1)]\;,\label{eq:phiT}\\
\phi_{P}^a(z,\omega^2)&=&\frac{3F_{P}^{\parallel}(\omega^2)}
{4\sqrt{2N_c}}(1-2z)[1+a_{2\rho}^a(10z^2-10z+1)]\;,\\
\phi_{}^v(z,\omega^2)&=&\frac{3F_{P}^{\parallel}(\omega^2)}
{8\sqrt{2N_c}}\bigg\{[1+(1-2z)^2]+a^v_{2\rho}[3(2z-1)^2-1]\bigg\}\;,\label{eq:phiv}
\end{eqnarray}
with the Gegenbauer coefficients $a_{2\rho}^{0,s,t}$ and $a_{2\phi}^{T,a,v}$, and two $P$-wave form factors $F_P^{\parallel}(\omega^2)$ and $F_P^{\perp}(\omega^2)$.
As mentioned above, the $P$-wave $\pi\pi$ DAs associated with the longitudinal polarization have been determined in a recent global analysis from the three-body decays $B_{(s)}\to P\rho\to P(\pi\pi)$, $P=\pi, K$, in the PQCD approach.
In this work, taking the four-body decay $B^0\to \rho^+\rho^-\to (\pi^0\pi^+)(\pi^0\pi^-)$ as an example,
we find that the calculated longitudinal branching ratio ${\cal B}(B^0\to \rho^+\rho^-\to (\pi^0\pi^+)(\pi^0\pi^-))$ shows a strong dependence on the variation of the Gegenbauer parameters,
which can be seen easily in Fig~\ref{a0st}.
Thus we will update the fitting results of the moments $a^{0,s,t}_{2\rho}$ in Eqs.~(\ref{eq:phi0})-(\ref{eq:phit}) by considering the  measured branching ratios in three-body and four-body charmless hadronic $B$ meson decays,
which will be expressed in detail in the following section.
Since the experimental measurements for the transverse components of the considered $B\to \rho\rho\to (\pi\pi)(\pi\pi)$ are not available,
the moments $a_{2\rho}^{T,a,v}$ in Eqs.~(\ref{eq:phiT})-(\ref{eq:phiv}) can not be fitted from the global analysis at present.
Their values are adopted the same  as those determined in Ref.~\cite{prd98-113003}: $a_{2\rho}^{T}=0.5\pm 0.5, a_{2\rho}^{a}=0.4\pm 0.4$ and $a_{2\rho}^{v}=-0.5\pm 0.5$.

According to the Watson theorem~\cite{pr88-1163}, the elastic rescattering effects in a final-state meson pair can be absorbed into the time-like form factor
$F(\omega^2)$.
The contribution from the wide $\rho$ resonance is usually parameterized as the Gounaris-Sakurai (GS)
model~\cite{prl21-244} based on the BW function~\cite{BW-model} in the experimental analysis of multi-body hadronic $B$ meson decays,
in which way the observed structures beyond the $\rho$ resonance in terms of higher mass isovector vector mesons can be well interpreted.
By taking the $\rho-\omega$ interference and the excited states into account,
the form factor $F_P^{\parallel}(\omega^2)$ can be described as~\cite{prd86-032013}
\begin{eqnarray}
F_P^{\parallel}(\omega^2)= \left [ {\rm GS}_\rho(s,m_{\rho},\Gamma_{\rho})
\frac{1+c_{\omega} {\rm BW}_{\omega}(s,m_{\omega},\Gamma_{\omega})}{1+c_{\omega}}
+\sum\limits_i c_i {\rm GS}_i(s,m_i,\Gamma_i)\right] \left[ 1+\sum\limits_i c_i\right]^{-1}\;,
\label{GS}
\end{eqnarray}
where $s=\omega^2$ is the two-pion invariant mass squared, $i=(\rho^{\prime}(1450), \rho^{\prime \prime}(1700), \rho^{\prime \prime \prime}(2254))$,
$\Gamma_{\rho,\omega,i}$ is the decay width for the relevant resonance, $m_{\rho,\omega,i}$ are the masses of the corresponding mesons, respectively.
The explicit expressions of the function ${\rm GS}_\rho(s,m_{\rho},\Gamma_{\rho})$ are described as follows~\cite{BW-model}
\begin{equation}
{\rm GS}_\rho(s, m_\rho, \Gamma_\rho) =
\frac{m_\rho^2 [ 1 + d(m_\rho) \Gamma_\rho/m_\rho ] }{m_\rho^2 - s + f(s, m_\rho, \Gamma_\rho)
- i m_\rho \Gamma (s, m_\rho, \Gamma_\rho)}~,
\end{equation}
with the functions
\begin{eqnarray}
\Gamma (s, m_\rho, \Gamma_\rho) &=& \Gamma_\rho  \frac{s}{m_\rho^2}
\left( \frac{\beta_\pi (s) }{ \beta_\pi (m_\rho^2) } \right) ^3~,\non
d(m) &=& \frac{3}{\pi} \frac{m_\pi^2}{k^2(m^2)} \ln \left( \frac{m+2 k(m^2)}{2 m_\pi} \right)
   + \frac{m}{2\pi  k(m^2)}
   - \frac{m_\pi^2  m}{\pi k^3(m^2)}~,\non
f(s, m, \Gamma) &=& \frac{\Gamma  m^2}{k^3(m^2)} \left[ k^2(s) [ h(s)-h(m^2) ]
+ (m^2-s) k^2(m^2)  h'(m^2)\right]~,\non
k(s) &=& \frac{1}{2} \sqrt{s}  \beta_\pi (s)~,\non
h(s) &=& \frac{2}{\pi}  \frac{k(s)}{\sqrt{s}}  \ln \left( \frac{\sqrt{s}+2 k(s)}{2 m_\pi} \right),
\end{eqnarray}
where $\beta_\pi (s) = \sqrt{1 - 4m_\pi^2/s}$.
Due to the limited studies on the form factor $F_P^{\perp}(\omega^2)$, we use the two decay
constants $f_{\rho}^{(T)}$ of the intermediate particle to determine the ratio $F_P^{\perp}(\omega^2)/F_P^{\parallel}(\omega^2)\approx (f_{\rho}^T/f_{\rho})$.

For the scalar resonance $f_0(980)$, we adopt the Flatt\'e parametrization where the resulting line shape is above and below the threshold of the intermediate particle~\cite{plb63-228}.
However, the Flatt\'e parametrization shows a scaling invariance and does not allow for an extraction of individual partial decay widths when the coupling of a resonance to the channel opening nearby is very strong.
Here we take the modified Flatt\'e model suggested by D.V.~Bugg~\cite{prd78-074023} following the LHCb Collaboration~\cite{prd89-092006,prd90-012003},
\begin{eqnarray}
F_S(\omega^2)=\frac{m_{f_0(980)}^2}{m_{f_0(980)}^2-\omega^2-im_{f_0(980)}(g_{\pi\pi}\rho_{\pi\pi}+g_{KK}\rho_{KK}F^2_{KK})}\;.
\end{eqnarray}
The coupling constants $g_{\pi\pi}=0.167$ GeV and $g_{KK}=3.47g_{\pi\pi}$~\cite{prd89-092006,prd90-012003}
describe the $f_0(980)$ decay into the final states $\pi^+\pi^-$ and $K^+K^-$, respectively.
The exponential factor $F_{KK}=e^{-\alpha q^2_{K}}$   is introduced above the ${\bar K} K$ threshold
to reduce the $\rho_{KK}$ factor as invariant mass increases,
where $q_K$ is the momentum of the kaon in the ${\bar K} K$ rest frame and $\alpha=2.0\pm 0.25$ ${\rm GeV}^{-2}$~\cite{prd78-074023,prd89-092006}.
The phase space factors $\rho_{\pi\pi}$ and $\rho_{KK}$ read
as~\cite{prd87-052001,prd89-092006,plb63-228}
\begin{eqnarray}
\rho_{\pi\pi}=\frac23\sqrt{1-\frac{4m^2_{\pi^\pm}}{\omega^2}}
 +\frac13\sqrt{1-\frac{4m^2_{\pi^0}}{\omega^2}},\quad
\rho_{KK}=\frac12\sqrt{1-\frac{4m^2_{K^\pm}}{\omega^2}}
 +\frac12\sqrt{1-\frac{4m^2_{K^0}}{\omega^2}}.
\end{eqnarray}

\subsection{Helicity amplitudes}
It is known that the phase space of the four-body $B$ meson decay depends on  five kinematic variables:
three helicity angles  $\theta_{1(2)}, \varphi$ shown in Fig.~\ref{fig1} and two invariant masses $\omega_{1,2}$,
where the $\theta_{1(2)}$ is the angle between the $\pi^-(\pi^+)$ direction in the $\rho^-(\rho^+)$ rest frame
and the $\rho^-(\rho^+)$ direction in the $B^0$ rest frame, and $\varphi$ is the angle between the $\rho^+$ and $\rho^-$ meson decay planes.
In the $B_{(s)}$ meson rest frame,
the fivefold differential decay rate for the $B_{(s)}\rightarrow (\pi\pi)(\pi\pi)$ can be expressed as
\begin{eqnarray}\label{eq:decayrate}
\frac{d^5\mathcal{B}}{d\theta_1d\theta_2d\varphi d\omega_{1}d\omega_{2}}
=\frac{\tau_{B_{(s)}} k(\omega_1)k(\omega_2)k(\omega_1,\omega_2)}{16(2\pi)^6m_{B_{(s)}}^2} |A|^2, 
\end{eqnarray}
with the $B_{(s)}$ meson lifetime $\tau_{B_{(s)}}$,
and
\begin{eqnarray}
k(\omega_1,\omega_2)&=&\frac{\sqrt{[m_{B_{(s)}}^2-(\omega_1+\omega_2)^2][m_{B_{(s)}}^2-(\omega_1-\omega_2)^2]}}{2m_{B_{(s)}}},
\end{eqnarray}
is the momentum of the $\pi\pi$ pair in the $B_{(s)}$ meson rest frame.
The explicit expression of kinematic variables $k(\omega)$ is defined in the $\pi\pi$ center-of-mass frame
\begin{eqnarray}
k(\omega)=\frac{\sqrt{\lambda(\omega^2,m_{\pi}^2,m_{\pi}^2)}}{2\omega},
\end{eqnarray}
with the K$\ddot{a}$ll$\acute{e}$n function $\lambda (a,b,c)= a^2+b^2+c^2-2(ab+ac+bc)$ and $m_{\pi}$ being the final state mass.

One can find that the four-body phase space has been performed in the studies of the
$K\rightarrow \pi\pi l\nu$ decay~\cite{pr168-1858},
the semileptonic $\bar{B}\rightarrow D(D^*)\pi l \nu$ decays~\cite{prd48-3204},
semileptonic baryonic decays~\cite{prd85-094019,plb780-100},
as well as the four-body baryonic decays~\cite{plb770-348}.
One can confirm that Eq.~(\ref{eq:decayrate}) is equivalent to those in Refs.~\cite{prd85-094019,plb770-348} through appropriate variable changes.
Replacing the helicity angle $\theta$ by the meson momentum fraction $\zeta$ via Eq.~(\ref{eq:cos}),
the Eq.~(\ref{eq:decayrate}) is turned into
\begin{eqnarray}\label{eq:decayrate1}
\frac{d^5\mathcal{B}}{d\zeta_1d\zeta_2d \omega_1d \omega_2d\varphi}=
\frac{\tau_{B_{(s)}} k(\omega_1)k(\omega_2)k(\omega_1,\omega_2)}{4(2\pi)^6m_{B_{(s)}}^2\sqrt{1+4\alpha_1}\sqrt{1+4\alpha_2}}|A|^2.
\end{eqnarray}

For four-body $B_{(s)}\to (\pi\pi)(\pi\pi)$ decays, besides the dominant contribution from the vector resonance $\rho$,
the scalar $f_0$ resonance can also contribute in the selected $\pi\pi$ invariant mass regions.
Thereby, one can decompose the decay amplitudes into six helicity components $A_h$ with $h=VV$ (3), $VS$ (2), and $SS$,
where $V$ and $S$ denote the vector and scalar mesons, respectively.
The first three are commonly referred to as the $P$-wave amplitudes,
where both final state $\pi\pi$ pairs come from intermediate vector mesons.
In the transversity basis, a $P$-wave decay amplitude can be decomposed into three components:
$A_0$, for which the polarizations of the final-state vector mesons are
longitudinal to their momenta, and $A_{\parallel}$ ($A_\perp$),
for which the polarizations are transverse to the momenta and parallel (perpendicular) to each other.
As the $S$-wave $\pi\pi$ pair can arise from $R_1$ or $R_2$ labelled in Fig.~\ref{fig2}(a),
the corresponding single $S$-wave amplitude is denoted as $A_{VS} (A_{SV})$.
The double $S$-wave amplitude $A_{SS}$ is associated with the final state,
in which both $\pi\pi$ pairs are produced in the $S$ wave.
These helicity amplitudes are described in detail as follows
\begin{eqnarray}
A_{VV}&:& B_{(s)} \rightarrow \rho\rho \to (\pi\pi) (\pi\pi), \nonumber\\
A_{VS}&:& B_{(s)} \rightarrow  \rho f_0 \to (\pi\pi) (\pi\pi) ,\nonumber\\
A_{SS}&:& B_{(s)} \rightarrow  f_0 f_0 \to (\pi\pi) (\pi\pi) .
\end{eqnarray}
Including the $\zeta_{1,2}$ and azimuth-angle dependencies,
the total decay amplitude in Eq.~(\ref{eq:decayrate1}) can be written as a coherent sum of the $S$-, $P$-, and double $S$-wave components,
\begin{eqnarray}\label{eq:allampli}
A&=&A_0\frac{2\zeta_1-1}{\sqrt{1+4\alpha_1}}\frac{2\zeta_2-1}{\sqrt{1+4\alpha_2}}
+A_{\parallel}2\sqrt{2}\sqrt{\frac{\zeta_1(1-\zeta_1)+\alpha_1}{1+4\alpha_1}}
\sqrt{\frac{\zeta_2(1-\zeta_2)+\alpha_2}{1+4\alpha_2}}\cos\varphi \nonumber\\
&&
+i A_{\perp}2\sqrt{2}\sqrt{\frac{\zeta_1(1-\zeta_1)+\alpha_1}{1+4\alpha_1}}
\sqrt{\frac{\zeta_2(1-\zeta_2)+\alpha_2}{1+4\alpha_2}}\sin\varphi\non
&& +A_{VS}\frac{2\zeta_1-1}{\sqrt{1+4\alpha_1}}+A_{SV}\frac{2\zeta_2-1}{\sqrt{1+4\alpha_2}}+A_{SS}.
\end{eqnarray}

\section{Numerical Analysis}\label{sec:3}\label{sec3}
In this section,
we firstly explain how to perform the global fit of the Gegenbauer moments $a^{0,s,t}_{2\rho}$ in Eqs.~(\ref{eq:phi0})-(\ref{eq:phit})
to measured branching ratios  in three-body and four-body  charmless hadronic $B$ meson decays in the PQCD approach.
With the determined Gegenbauer moments,
the branching rations ($\cal B$), polarization fractions $f_{\lambda}$, direct $CP$ asymmetries,
along with the TPAs for the concerned four-body decays have been examined.
Other related input parameters,
including the meson masses, Wolfenstein parameters, the lifetimes~\cite{pdg2020}, as well as the decay constants~\cite{prd76-074018,2105-03899},
are also listed in Table~\ref{tab:constant}.

\begin{table}
\caption{The decay constants are taken from Refs.~\cite{prd76-074018,2105-03899}.
Other parameters are from PDG 2022~\cite{pdg2020}. }
\label{tab:constant}
\centering
\begin{tabular*}{14.5cm}{@{\extracolsep{\fill}}lllll}
  \hline\hline
\text{Mass(\text{GeV})}
& $m_{B_s}=5.37$  & $m_B=5.28$  &$m_{\pi^{\pm}}=0.140$ &$m_{\pi^{0}}=0.135$\\[1ex]
\text{Wolfenstein parameters}
& $\lambda=0.22650$  & $A=0.790$  &$\bar{\rho}=0.141$ & $\bar{\eta}=0.357$ \\[1ex]
\text{Decay constants (GeV)}
& $f_{B_s}=0.23$ & $f_{B}=0.21$  &$f_{\rho(770)}=0.216$ &$f_{\rho(1020)}^T=0.184$ \\[1ex]
\text{Lifetime (ps)}
& $\tau_{B_s}=1.520$ & $\tau_{B^0}=1.519$ & $\tau_{B^{\pm}}=1.638$\\[1ex]
\hline\hline
\end{tabular*}
\end{table}

\subsection{Global fit}
\begin{table}[htbp!]
	\centering
	\caption{Experimental data for branching ratios and polarization fractions \cite{pdg2020}, and the theoretical
	results derived from the fitted Gegenbauer moments in Eq.~(\ref{eq:phinew}).
The theoretical uncertainties are attributed to the variations of the shape parameter $\omega_{B_{(s)}}$ in
the $B_{(s)}$ meson DA, of the Gegenbauer moments in various twist DAs of the $\pi\pi$ pair,
and of the hard scale $t$ and the QCD scale $\Lambda_{\rm QCD}$, but added in quadrature.}
\begin{ruledtabular}
\begin{threeparttable}
    \setlength{\tabcolsep}{1mm}{
	\begin{tabular}{lccccccc}
		channel        &\multicolumn{3}{c}{data ${\cal B}(10^{-6})$ }              &\multicolumn{3}{c}{fit ${\cal B}(10^{-6})$}
  \cr \hline
$B^+ \to K^+\rho^0\to K^+(\pi^+\pi^-)$         &      & $3.7\pm 0.5$ &                      &     &$3.40^{+1.76}_{-1.03}$&   \\
$B^+ \to K^0\rho^+\to K^0(\pi^+ \pi^0)$          &      & $7.3^{+1.0}_{-1.2}$ &                     &     &$7.26^{+4.09}_{-2.34}$   &   \\
$B^+ \to \pi^+\rho^0\to \pi^+(\pi^+\pi^-)$         &      & $8.3\pm 1.2$ &                   &     &$6.74^{+1.94}_{-1.42}$   &\\
$B^+ \to \pi^0\rho^+\to \pi^0(\pi^+ \pi^0)$          &      & $10.9\pm 1.4$ &                 &     &$11.70^{+4.26}_{-3.05}$&   \\
$B^0 \to K^0\rho^0\to K^0(\pi^+\pi^-)$         &      & $3.4\pm 1.1$ &                      &     &$4.00^{+1.34}_{-1.03}$  &   \\
$B^0 \to K^+\rho^-\to K^+(\pi^- \pi^0)$         &      & $7.0\pm 0.9$ &                      &     &$9.29^{+4.77}_{-2.51}$     &   \\
$B^0 \to \pi^{+}\rho^{-}\to \pi^{+}(\pi^- \pi^0)$          &      & $23.0\pm 2.3$~\tnote{1} &           &     &$10.72^{+3.83}_{-2.81}$&   \\
$B^0 \to \pi^{-}\rho^{+}\to \pi^{-}(\pi^+ \pi^0)$          &      & $23.0\pm 2.3$~\tnote{1} &           &     &$21.79^{+8.39}_{-5.35}$&   \\
\hline
\multirow{2}{*}{channel}                &\multicolumn{3}{c}{data}                       &\multicolumn{3}{c}{fit}\cr\cline{2-7}
	                                     &${\cal B}(10^{-6})$      &   & $f_{0}(\%)$      &${\cal B}(10^{-6})$       &  & $f_{0}(\%)$\cr \hline
$B^0 \to \rho^+\rho^-\to(\pi^+\pi^0)(\pi^-\pi^0)$             &$27.7\pm 1.9$      &     & $99.0^{+2.1}_{-1.9}$          &$27.34^{+10.44}_{-7.73}$  &  & $93.8^{+2.5}_{-3.6}$    \\
\end{tabular}}
\begin{tablenotes}
\item $\tnote{1}$ Sum of two branching ratios, ${\cal B}(B\to f) + {\cal B}(B\to \bar{f})$.
\end{tablenotes}
\end{threeparttable}
\end{ruledtabular}
\label{tab:brfour}
\end{table}
The two-pion DAs given in Eqs.~(\ref{eq:phi0})-(\ref{eq:phit}) suggest that
the longitudinal amplitude $A^{L}$  of the four-body decay $B\to \rho\rho \to (\pi\pi)(\pi\pi)$
can be expanded in terms of the Gegenbauer moments.
We then  decompose the squared amplitude $|A^{L}|^2$
into the linear combinations of the Gegenbauer moments $a^{0,s,t}_{2\rho}$ together with the corresponding products:
\begin{eqnarray}
|A^{L}|^2&=&M^L_0+a^0_{2\rho}M^L_1+a^s_{2\rho}M^L_2+a^t_{2\rho}M^L_3+(a^0_{2\rho})^2M^L_4+(a^s_{2\rho})^2M^L_5+(a^t_{2\rho})^2M^L_6+a^0_{2\rho}a^s_{2\rho}M^L_7\nonumber\\
&+&a^0_{2\rho}a^t_{2\rho}M^L_8+a^s_{2\rho}a^t_{2\rho}M^L_9+a^0_{2\rho}a^s_{2\rho}a^t_{2\rho}M^L_{10}+(a^0_{2\rho})^3M^L_{11}+(a^s_{2\rho})^3M^L_{12}
+(a^t_{2\rho})^3M^L_{13}+a^s_{2\rho}(a^0_{2\rho})^2M^L_{14}\non
&+&a^t_{2\rho}(a^0_{2\rho})^2M^L_{15}+
a^0_{2\rho}(a^s_{2\rho})^2M^L_{16}+a^t_{2\rho}(a^s_{2\rho})^2M^L_{17}+a^0_{2\rho}(a^t_{2\rho})^2M^L_{18}+a^s_{2\rho}(a^t_{2\rho})^2M^L_{19}+(a^0_{2\rho})^4M^L_{20}\non
&+&(a^s_{2\rho})^4M^L_{21}+(a^t_{2\rho})^4M^L_{22}
+a^0_{2\rho}(a^s_{2\rho})^3M^L_{23}+a^t_{2\rho}(a^s_{2\rho})^3M^L_{24}+a^0_{2\rho}(a^t_{2\rho})^3M^L_{25}+a^s_{2\rho}(a^t_{2\rho})^3M^L_{26}\non
&+&a^s_{2\rho}(a^0_{2\rho})^3M^L_{27}+a^t_{2\rho}(a^0_{2\rho})^3M^L_{28}+
(a^0_{2\rho}a^s_{2\rho})^2M^L_{29}+(a^0_{2\rho}a^t_{2\rho})^2M^L_{30}+(a^s_{2\rho}a^t_{2\rho})^2M^L_{31}\non
&+&a^0_{2\rho}a^s_{2\rho}(a^t_{2\rho})^2M^L_{32}+a^0_{2\rho}a^t_{2\rho}(a^s_{2\rho})^2M^L_{33}+a^t_{2\rho}a^s_{2\rho}(a^0_{2\rho})^2M^L_{34}.\label{eq:ampL}
\end{eqnarray}
While the parametrization formulas of the squared amplitudes $|A|^2$ for three-body decays $B_{(s)}\to (\pi,K)\rho\to (\pi,K)\pi\pi$
have been presented in Ref.~\cite{2105-03899},
\begin{eqnarray}
|A|^2 &= & M_{0\rho}+a^{0}_{2\rho}M_{1}+(a^{0}_{2\rho})^2M_{2}+a^{s}_{2\rho}M_{3}+(a^{s}_{2\rho})^2M_{4}\non
           &+&      a^{t}_{2\rho}M_{5}+(a^{t}_{2\rho})^2 M_{6}+(a^{0}_{2\rho})(a^{s}_{2\rho})M_{7}+
                 (a^{0}_{2\rho})(a^{t}_{2\rho})M_{8}+(a^{s}_{2\rho})(a^{t}_{2\rho})M_{9}.\label{eq:amp3}
\end{eqnarray}
The coefficients $M$ in above equations, involving only the Gegenbauer polynomials, are perturbatively calculable in the PQCD approach
and therefore can be used to establish the database for the global fit.

Similar to the proposal in Ref.~\cite{2012-15074},
we can fit the PQCD factorization formulas with the database constructed in Eqs.~(\ref{eq:ampL}) and (\ref{eq:amp3})
to the nine pieces of $B^{+}\to K^+\rho^{0}\to K^+(\pi\pi)$, $B^{0}\to K^+\rho^{-}\to K^+(\pi\pi)$,
$B^{+}\to K^0\rho^{+}\to K^0(\pi\pi)$, $B^{0}\to K^0\rho^{0}\to K^0(\pi\pi)$
$B^{+}\to \pi^+\rho^{0}\to \pi^+(\pi\pi)$, $B^{+}\to\pi^0\rho^{+}\to \pi^0(\pi\pi)$, $B^{0}\to \pi^{\pm}\rho^{\mp}\to\pi^{\pm} (\pi\pi)$
and $B^0\to \rho^+\rho^- \to (\pi\pi)(\pi\pi)$ data,
including eight branching ratios and one polarization fraction.
We adopt the standard nonlinear least-$\chi^2$ (lsq) method~\cite{Peter:2020}, in which
the $\chi^2$ function is defined for $n$ pieces of experimental data
$v_i\pm \delta v_i$ with the errors  $\delta v_i$ and the fitted corresponding theoretical
values $v^{\rm{th}}_i$ as
\begin{eqnarray} \label{eq:fit}
	\chi^2= \sum_{i=1}^{n}  \Big(\frac {v_i - v^{\rm{th}}_i}{\delta v_i}\Big)^2.
\end{eqnarray}
We usually should include maximal amount of data in the fit in order to minimize statistical uncertainties.
But
those measurements with significance lower than 3$\sigma$ do not impose stringent constraints,
and need not be taken into account in principle.
What's more,
the experimental data of those decay modes,
such as $B^{0} \to \rho^{0}\rho^{0}\to (\pi^+\pi^-)(\pi^+\pi^-)$ and $B^{+} \to \rho^{+}\rho^{0}\to (\pi^+\pi^0)(\pi^+\pi^-)$,
which receive substantial next-to-leading order (NLO) corrections, are also excluded,
even though they may have higher precision.

The improved fitting results of the Gegenbauer moments $a^{0,s,t}_{2\rho}$ with $\chi^2 / d.o.f.=2.1$ are summarized as follows:
\begin{eqnarray}\label{eq:phinew}
a^0_{2\rho}=0.39\pm0.11, \quad\quad  a^s_{2\rho}=-0.34\pm0.26,\quad\quad a^t_{2\rho}=-0.13\pm0.04,
\end{eqnarray}
whose errors mainly arise from the experimental uncertainties.
For comparison, the updated PQCD predictions of the branching ratios are also listed in Table~\ref{tab:brfour} and basically agree with the data within the  large uncertainties.
It should also be noted that,
apart from the inclusion of the new four-body decay $B^0\to \rho^+\rho^- \to (\pi^+\pi^0)(\pi^-\pi^0)$,
the $\pi\pi$ invariant mass rang of the three-body channels $B_{(s)} \to (\pi, K)\rho \to (\pi, K)\pi\pi$
have been modified to be $0.3<m_{\pi\pi}<1.1$ GeV instead of $2m_{\pi}<m_{\pi\pi}<m_{B}-m_{\pi(K)}$ GeV in the present work.
The updated fitting results of the $a^0_{2\rho}=0.39\pm0.11, a^s_{2\rho}=-0.34\pm0.26, a^t_{2\rho}=-0.13\pm0.04$
could then be a bit different from the previous ones $a^{0}_{2\rho}=0.08\pm0.13, a^{s}_{2\rho}=-0.23\pm0.24, a^{t}_{2\rho}=-0.35\pm0.06$.

\subsection{Branching rations and polarization fractions of the two-body $B_{(s)} \to \rho\rho$ decays}
On basis of Eq.~(\ref{eq:decayrate1}), one can obtain the branching ratio as
\begin{eqnarray}\label{eq:brsss}
\mathcal{B}_h=\frac{\tau_{B_{(s)}}}{4(2\pi)^6m_{B_{(s)}}^2}\frac{2\pi}{9}C_h
\int d\omega_1d\omega_2k(\omega_1)k(\omega_2)k(\omega_1,\omega_2)|A_h|^2,
\end{eqnarray}
with the invariant masses $\omega_{1,2}$ integrating over the chosen $\pi\pi$ mass window.
The coefficients $C_h$ are the results of the integrations over $\zeta_1,\zeta_2,\varphi$ in terms of Eq.~(\ref{eq:brsss}) and listed as follows,
\begin{eqnarray}\label{eq:radii}
C_h=\left\{\begin{aligned}
&(1+4\alpha_1)(1+4\alpha_2), \quad\quad\quad  &h=0,\parallel,\perp, \\
&3(1+4\alpha_{1, 2}) ,\quad\quad\quad  &h=VS, SV,\\
&9, \quad\quad\quad  &h=SS. \\
\end{aligned}\right.
\end{eqnarray}
Combining Eq.~(\ref{eq:brsss}) with its counterpart of the corresponding $CP$-conjugated process,
the $CP$-averaged branching ratio of each component can be defined as below,
\begin{eqnarray}
\mathcal{B}_h^{\rm avg}=\frac{1}{2}(\mathcal{B}_h+\mathcal{\bar{B}}_h).
\end{eqnarray}
The polarization fractions $f_{\lambda}$ with $\lambda=0$, $\parallel$,
and $\perp$ in the $B_{(s)}\to VV$ decays are described as
\begin{eqnarray}\label{pol}
f_{\lambda}=\frac{\mathcal{B}_{\lambda}}{\mathcal{B}_0+\mathcal{B}_{\parallel}+\mathcal{B}_{\perp}},
\end{eqnarray}
satisfying  the normalization relation $f_0+f_{\parallel}+f_{\perp}=1$.

\begin{table}[!htbh]
\caption{Branching ratios and polarization fractions  of the two-body $B_{(s)}\rightarrow \rho \rho$ decays.
For a comparison, we also list the results from the previous PQCD \cite{prd91-054033},
QCDF \cite{prd80-114026}, SCET \cite{prd96-073004}, and FAT \cite{epjc77-333}.
The world averages of experimental data are taken from PDG 2022~\cite{pdg2020}.
The theoretical uncertainties are attributed to the variations of the shape parameter $\omega_{B_{(s)}}$ in
the $B_{(s)}$ meson DA, of the Gegenbauer moments in various twist DAs of the $\pi\pi$ pair,
and of the hard scale $t$ and the QCD scale $\Lambda_{\rm QCD}$.}
\label{tab:brtwo}
\begin{ruledtabular}  \begin{threeparttable}
\setlength{\tabcolsep}{1mm}{ \begin{tabular}[t]{lcccc }
Modes     & $\mathcal{B}(10^{-6})$       & $f_0(\%)$  & $f_\|(\%)$  & $f_\perp(\%)$
                   \\ \hline
$B^+\rightarrow\rho^+ \rho^0$    &$12.87_{-2.97-1.56-0.74}^{+4.07+1.59+0.91}$       &$98.13_{-0.01-0.73-0.17}^{+0.01+0.60+0.16}$
                                 &$0.82_{-0.00-0.36-0.05}^{+0.01+0.41+0.07}$     &$1.05_{-0.00-0.26-0.06}^{+0.01+0.32+0.09}$ \\
PQCD~\cite{prd91-054033}         &$13.5^{+5.1}_{-4.1}$      &$98\pm1$             & $\cdots$  &$0.46^{+0.08}_{-0.06}$\\
QCDF~\cite{prd80-114026}         &$20.0^{+4.5}_{-2.1}$     &$96\pm2$       & $\cdots$  &$\cdots$             \\
SCET~\cite{prd96-073004}         &$22.1\pm 3.7$              &$100$                & $\cdots$  &$\cdots$       \\
FAT~\cite{epjc77-333}            &$21.7\pm5.1$              &$95.5\pm1.1$             & $\cdots$  &$2.22\pm 0.64$           \\
Data~\cite{pdg2020}              &$24.0\pm 1.9$              &$95.0\pm 1.6$             & $\cdots$  &$\cdots$       \\\hline
$B^0\rightarrow\rho^+ \rho^-$    &$27.34_{-6.75-3.44-1.54}^{+9.57+3.61+2.09}$     &$93.82_{-0.23-3.54-0.23}^{+0.41+2.85+0.18}$
                                 &$2.56_{-0.10-1.41-0.69}^{+0.11+1.74+0.13}$     &$3.62_{-0.13-1.58-0.10}^{+0.14+1.91+0.12}$ \\
PQCD~\cite{prd91-054033}         &$26.0^{+10.3}_{-8.3}$      &$95\pm1$             & $\cdots$  &$2.42^{+0.21}_{-0.19}$\\
QCDF~\cite{prd80-114026}         &$25.5^{+1.9}_{-3.0}$       &$92^{+1}_{-3}$       & $\cdots$  &$\cdots$             \\
SCET~\cite{prd96-073004}         &$27.7\pm 4.1$              &$99.1\pm 0.3$                & $\cdots$  &$0.40\pm0.18$       \\
FAT~\cite{epjc77-333}            &$29.5\pm6.5$              &$92.6\pm1.6$             & $\cdots$  &$3.65\pm 0.91$           \\
Data~\cite{pdg2020}              &$27.7\pm1.9$              &$99.0^{+2.1}_{-1.9}$             & $\cdots$  &$\cdots$       \\\hline
$B^0\rightarrow\rho^0 \rho^0$    &$0.30_{-0.05-0.04-0.03 }^{+0.06+0.03+0.02}$       &$41.81_{-1.89-6.47-3.52}^{+2.96+4.58+4.92}$
&$26.38_{-0.58-2.23-1.95}^{+1.01+2.92+1.54}$     &$31.81_{-0.24-2.85-2,97}^{+1.08+3.10+1.98}$ \\
PQCD~\cite{prd91-054033}         &$0.27^{+0.12}_{-0.10}$      &$12^{+16}_{-2}$             & $\cdots$  &$45.9^{+1.1}_{-8.2}$\\
QCDF~\cite{prd80-114026}         &$0.9^{+1.9}_{-0.45}$     &$92^{+7}_{-37}$       & $\cdots$  &$\cdots$             \\
SCET~\cite{prd96-073004}         &$1.00\pm 0.29$              &$87\pm 5$                & $\cdots$  &$5.81\pm 2.84$       \\
FAT~\cite{epjc77-333}            &$0.94\pm 0.59$              &$81.7\pm10.8$             & $\cdots$  &$9.21\pm 5.50$           \\
Data~\cite{pdg2020}              &$0.96\pm0.15$              &$71^{+8}_{-9}$             & $\cdots$  &$\cdots$       \\\hline
$B_s^0\rightarrow\rho^+ \rho^-$    &$2.04_{-0.34-0.36-0.50 }^{+0.43+0.42+0.55}$       &$99.65_{-0.04-0.36-0.06}^{+0.03+0.27+0.05}$
                                   &$0.16_{-0.01-0.13-0.01}^{+0.01+0.12+0.01}$     &$0.19_{-0.03-0.16-0.04}^{+0.03+0.27+0.06}$ \\
PQCD~\cite{prd91-054033}         &$1.5^{+0.7}_{-0.6}$      &$\sim 1.0$             & $\cdots$  &$\sim 0.0$\\
QCDF~\cite{prd80-114026}         &$0.68^{+0.73}_{-0.53} $     &$\sim 1.0$       & $\cdots$  &$\cdots$             \\
FAT~\cite{epjc77-333}            &$0.10\pm 0.06$              &$\cdots$             & $\cdots$  &$\cdots$           \\\hline
$B_s^0\rightarrow\rho^0 \rho^0$    &$1.02_{-0.17-0.18-0.25 }^{+0.22+0.21+0.28}$       &$99.65_{-0.04-0.36-0.06}^{+0.03+0.27+0.05}$
                                   &$0.16_{-0.01-0.13-0.01}^{+0.01+0.12+0.01}$     &$0.19_{-0.03-0.16-0.04}^{+0.03+0.27+0.06}$ \\
PQCD~\cite{prd91-054033}         &$0.74^{+0.45}_{-0.28}$      &$\sim 1.0$             & $\cdots$  &$\sim 0.0$\\
QCDF~\cite{prd80-114026}         &$0.34^{+0.36}_{-0.26}$     &$\sim 1.0$       & $\cdots$  &$\cdots$             \\
FAT~\cite{epjc77-333}            &$0.05\pm 0.03$             &$\cdots$             & $\cdots$  &$\cdots$               \\
Data~\cite{pdg2020}              &$<320$              &$\cdots$             & $\cdots$  &$\cdots$       \\
\end{tabular}}
\end{threeparttable}
\end{ruledtabular}
\end{table}

According to the narrow width approximation equation
\beq
\label{2body}
\mathcal{B}(B_{(s)} \rightarrow  \rho\rho\rightarrow (\pi\pi)(\pi\pi))&\approx& \mathcal{B}(B_{(s)} \rightarrow \rho\rho)\times \mathcal{B}(\rho \rightarrow \pi\pi)
 \times \mathcal{B}(\rho \rightarrow \pi\pi),
\eeq
with $\mathcal{B}(\rho \rightarrow \pi\pi)=100\%$,
we extract the two-body branching ratios ${\cal B}(B_{(s)} \to \rho\rho)$ from the corresponding four-body $B_{(s)} \rightarrow \rho\rho \rightarrow (\pi\pi) (\pi\pi)$ decays and summarize them in Table~\ref{tab:brtwo},
in which the first quoted uncertainty is due to the shape parameters
$\omega_{B_{(s)}}=0.40\pm 0.04 (0.48\pm 0.048)$ in the initial $B_{(s)}$  meson DAs ,
the second uncertainty is caused by  the Gegenbauer moments in the $\pi\pi$ DAs shown in Eqs.~(\ref{eq:phis0}) and Eqs.~(\ref{eq:phi0})-(\ref{eq:phiv}),
and the last one comes from the variation of the hard scale $t$ from $0.75t$ to $1.25t$ together with the QCD scale
$\Lambda_{QCD}=0.25\pm 0.05$ GeV.
The polarization fractions of the two-body decays $B_{(s)} \to \rho\rho$ calculated in this work are also given in Table~\ref{tab:brtwo}.
For the sake of comparison,
we also present the numerical results from the previous  PQCD~\cite{prd91-054033}, QCDF~\cite{prd80-114026}, SCET~\cite{prd96-073004}, and FAT~\cite{epjc77-333},
while the experimental measurements for branching ratios, polarization fractions are taken from PDG 2022~\cite{pdg2020}.

It is seen that all of the theoretical predictions of ${\cal B}(B^0\to \rho^+\rho^-)$ are consistent with experimental measurements within errors.
While for $B^+\to \rho^+\rho^0$ and $B^0\to \rho^0\rho^0$ decays,
our predictions are $(12.87^{+4.46}_{-3.43})\times 10^{-6}$ and $(0.30^{+0.07}_{-0.06})\times 10^{-6}$ respectively,
which are a bit smaller than those of the QCDF~\cite{prd80-114026},  SCET~\cite{prd96-073004}
and FAT~\cite{epjc77-333} as well as the experimental data~\cite{pdg2020},
but close to the previous two-body PQCD results~\cite{prd91-054033}.
For $B^+\to \rho^+\rho^0$ decay, judging from the isospin triangle,
the relation of $2{\cal B}(B^+\to \rho^+\rho^0)={\cal B}(B^0\to \rho^+\rho^-)$ is expected
because of the quite small decay rate of $B^0\to \rho^0\rho^0$ (${\cal B}\sim 10^{-7}$).
On the experimental side, the two rates are basically equal to each other within errors,
which is puzzling and needs to be resolved in the future.
It is worthwhile to mention that large cancellations exists in the leading-order (LO) matrix elements for the color-suppressed ($``C"$) decay $B^0\to \rho^0\rho^0$,
which leads to the small branching fraction $(0.30^{+0.07}_{-0.06})\times 10^{-6}$.
As shown in Refs.~\cite{prd73-114014,cpc46-123103},
if the next-to-leading order (NLO) corrections are taken into account,
the two-body branching ratio ${\cal B}(B^0\to \rho^0\rho^0)$ could be enhanced a lot
and become comparable with the experimental measurement.
The gap between the PQCD calculation and the data can be solved effectively
when the NLO contributions are considered.
However, the NLO corrections to the four-body decays in the PQCD framework is a big task and goes beyond the scope of the present work.

The decays $B^0_{s}\to \rho^{+}\rho^{-}$ and $B^0_{s}\to \rho^{0}\rho^{0}$ can occur only via annihilation topology in the SM.
The predicted ${\cal B}(B^0_{s}\to \rho^{+}\rho^{-})=(2.04^{+0.81}_{-0.70})\times 10^{-6}$
and ${\cal B}(B^0_{s}\to \rho^{0}\rho^{0})=(1.02^{+0.41}_{-0.35})\times 10^{-6}$ are about twice larger
than the QCDF results ${\cal B}(B^0_{s}\to \rho^{+}\rho^{-})=(0.68^{+0.73}_{-0.53})\times 10^{-6}$
and ${\cal B}(B^0_{s}\to \rho^{0}\rho^{0})=(0.34^{+0.36}_{-0.26})\times 10^{-6}$.
It is known the annihilation diagrams in the QCDF and SCET framework have to be fitted from the experimental measurements due to the endpoint singularities.
In Ref.~\cite{npb990-116175},
the authors have studied systematically weak annihilation $B$ decays and proposed a new way to treat the end-point contribution.
While in Ref.~\cite{prd96-073004}, the contributions from the annihilation diagrams that belong to next-to-leading power in the SCET scheme has been neglected.
The very distinct results in the FAT method from those in the PQCD and QCDF approaches need to be further examined by the coming measurements from LHCb and Belle-II experiments.
We remark that, since  the pure annihilation $B_s^0 \to \pi^+\pi^-$ decay rate has already been confirmed by the CDF~\cite{prl108-211803} and LHCb ~\cite{prl118-081801} Collaborations with high precision,
the large branching fractions of the $B^0_{s}\to \rho^{+}\rho^{-}$ and $B^0_{s}\to \rho^{0}\rho^{0}$ decay modes ( $\sim 10^{-6}$), are expected to be verified soon.

For the charmless $B\to VV$ decays, it is expected that three helicity amplitudes $A_{0,\|,\bot}$ obey the following hierarchy pattern according to the naive counting rules~\cite{plb662-63},
\begin{eqnarray}
A_0:A_{+}:A_{-}=1:\frac{\Lambda_{\rm QCD}}{m_b}:(\frac{\Lambda_{\rm QCD}}{m_b})^2,
\end{eqnarray}
with $A_{\pm}=(A_{\|}\pm A_{\bot})/\sqrt{2}$.
The above hierarchy relation satisfies the expectation in the
factorization assumption that the longitudinal polarization
should dominate based on the quark helicity analysis~\cite{zpc1-269,prd64-117503}.
However,
experimental measurements of the low longitudinal polarization fractions (of order $50\%$)
for penguin-dominated decays, like $B\to K^*\rho$, $B\to K^*\phi$,
and $B^0_s\to \phi\phi$~\cite{prd85-072005,jhep05-026,prl91-201801,prd78-092008,prl107-261802,plb713-369} modes,
 are in conflict with this expectation,
which poses an interesting challenge for the theory.
Several strategies have been proposed  to interpret the large transverse polarizations within or beyond the SM~\cite{npb774-64,prd71-054025,prd70-054015,Cheng:2008gxa,Grossman:2003qi,Das:2004hq,Chen:2005mka,Yang:2004pm,
Kagan:2004uw,Beneke:2005we,Datta:2007qb,Colangelo:2004rd,Ladisa:2004bp,
Cheng:2004ru,
Bobeth:2014rra,Cheng:2010yd,Chen:2007qj,Chen:2005cx,Faessler:2007br,Chen:2006vs,Huang:2005qb,Baek:2005jk,Yang:2005tv,Alvarez:2004ci}.

From the numerical results presented in Table~\ref{tab:brtwo},
one can observe that the longitudinal polarizations of the decays $B^0\to \rho^+\rho^-$ and $B^+\to \rho^+\rho^0$ can be as large as $90\%$,
which are basically consistent  with those from QCDF~\cite{prd80-114026}, SCET~\cite{prd96-073004}, FAT~\cite{epjc77-333},
and the current experimental data~\cite{pdg2020}.
The dominant contributions of the $B^0\to \rho^+\rho^-$ and $B^+\to \rho^+\rho^0$ decays
are from the terms $(C_1/3+C_2 )F^{LL, i}_{e\rho}$ ($i=0,\parallel,\perp$),
induced by the tree operators $O_{1}$ and $O_{2}$.
The transverse amplitudes $F_{e\rho}^{LL, \parallel}$ and $F_{e\rho}^{LL, \perp}$ are always power suppressed
relative to the longitudinal ones $F_{e\rho}^{LL, 0}$, which leads to $f_0\sim 90\%$.

The calculated longitudinal fraction $f_0$  of the $``C"$-type decay $B^0\to \rho^0\rho^0$
are around $40\%$,
which is much smaller than that predicted in the QCDF~\cite{prd80-114026}, SCET~\cite{prd96-073004}, FAT~\cite{epjc77-333} and the data~\cite{pdg2020}.
For the color-suppressed decay,
the longitudinal polarization contributions from two hard-scattering emission diagrams largely cancel against each other in the PQCD approach.
While the chiral enhanced annihilation diagrams induced by the QCD operator $O_6$,
together with the hard scattering emission diagrams can provide a significant transverse contribution.
As a result, the transverse polarization fraction is almost comparable with the longitudinal one.
Similar to the situation of the branching fraction ${\cal B}(B^0\to \rho^0\rho^0)$,
the calculated $f_0(B^0\to \rho^0\rho^0)$ can also receive substantial NLO contributions~\cite{prd73-114014,cpc46-123103}.
As already stressed above,
it is not practical  for us to analyze the four-body decays $B\to \rho\rho\to (\pi\pi)(\pi\pi)$ in the NLO accuracy at present,
which should  be left for the future studies.

For the two pure-annihilation decays $B^0_s\to \rho^0\rho^0$ and $B^0_s\to \rho^+\rho^-$,
the longitudinal polarizations dominate the contributions, $f_0\approx 1$,
which also agree well with the previous two-body PQCD \cite{prd91-054033} and QCDF \cite{prd80-114026} analysis.
It is known that there is actually a big cancellation between the two factorizable diagrams Figs.~\ref{fig2}(e) and ~\ref{fig2}(f),
especially when two final state mesons are identical.
The quantities $F^{LL,0,\parallel}_{a\rho}$ and $F^{LR,0,\parallel}_{a\rho}$ in the $B^0_s\to \rho^0\rho^0$ and $B^0_s\to \rho^+\rho^-$
decays are equal to zero because of the current conservation.
The only left amplitudes $F^{LL,\perp}_{a\rho}$ and $F^{LR,\perp}_{a\rho}$
for the factorizable diagrams are highly suppressed by the factor $(\omega_{\pi\pi}/M_{B_s})^2\approx (m_\rho/M_{B_s})^2 \approx 0.02$.
For the non-factorizable annihilation diagrams Figs.~\ref{fig2}(g) and ~\ref{fig2}(h),
the longitudinal components $M^{LL,0}_{a\rho}$ and $M^{SP,0}_{a\rho}$ will give the leading and dominant contributions,
and other terms related to both parallel and perpendicular components are also suppressed by $(\omega_{\pi\pi}/M_{B_s})^2\approx (m_\rho/M_{B_s})^2 \approx 0.02$.
Thus, the total transverse polarization contributions of the decays $B^0_s\to \rho^0\rho^0$ and $B^0_s\to \rho^+\rho^-$ are indeed rather small,
resulting in  $f_0\approx 1$ as shown in Table~\ref{tab:brtwo}.

\subsection{$S$-wave fractions}
As we know, the identification of scalar resonances is a long-standing puzzle compared with the well studied vector resonances.
It is difficult to deal with scalar resonances since some of them have wide decay widths, which cause a strong overlap between resonances and background.
Furthermore, the underlying structure of scalar meson is not well established in the theory side (for a review, see Ref.~\cite{pdg2020}).
At present,
two main different scenarios~\cite{prd73-014017}, the so-called scenario I (S-I) and scenario II (S-II),
have been proposed to interpret the scalar mesons.
For instance,
the $f_0$ is viewed as the conventional two-quark $q{\bar q}$ state in S-I,
while as the four-quark  bound state in S-II.
However,
the $f_0$ in S-II  described by the four-quark state is too complicated to be studied in a factorization approach.
In order to give quantitative predictions,
we will only work in the S-I for $f_0$.
Moreover,
similar to the situation of the $\eta-\eta^{\prime}$ mixing,
the experimental measurements of the decays $D_s^+\to f_0\pi$, $\phi\to f_0\gamma$, $J/\psi \to f_0 \omega$, $J/\psi \to f_0 \phi$, etc.~\cite{pdg2020}
clearly show that the scalar resonance $f_0$ has both nonstrange and strange quark components,
and can be regarded as a mixture of $s {\bar s}$ and $n{\bar n}=(u {\bar u}+d {\bar d})/\sqrt{2}$,
\begin{eqnarray}
|f_0\rangle=|n {\bar n}\rangle {\rm sin} \theta+|s {\bar s}\rangle{\rm cos}\theta.\label{f0mix}
\end{eqnarray}
The mixing angle $\theta$ in the above equation has not been determined precisely by current experimental measurements,
and is suggested to be in the wide ranges of $25^{\circ} <\theta < 40^{\circ}$ and $140^{\circ} <\theta < 165^{\circ}$~\cite{prd67-034024,pan65-497,plb609-291}.
For simplicity,
we will adopt the value $\theta=145^{\circ}$~\cite{epjc82-59,prd103-113005} in our calculation.
It should be noted that
the scalar meson $f_0(980)$ is usually taken into account in the pure $s {\bar s}$ density operator in our previous works~\cite{zjhep,Li:2021qiw,prd105-053002,prd105-093001}.
Besides,
different kinds of theoretical approaches have also been applied to study the
$B_{(s)}$ meson decays involving $f_0(980)$ in the final states with the assumption that $f_0(980)$ is a pure ${\bar s}s$ state~\cite{prd81-074001,prd83-094027,epjc80-554}.
However,
taking the  measured decay $B^0\to \rho^0f_0$ as an example,
we find numerically that the contribution from the $f_0=(u {\bar u}+d {\bar d})/\sqrt{2}$ component is dominant,
as can be seen easily in Table~\ref{tab:fnfs}.
Therefore,
the mixing effect of the $f_0(980)$ meson shown in Eq.~(\ref{f0mix}) should be involved in the present work.
\begin{table}[!htbh]
\caption{Branching ratio of the considered $S$-wave decay
 $B^0\to \rho^0 f_0\to (\pi^+\pi^-)(\pi^+\pi^-)$.
The subscripts $f_n$ and $f_s$ represent the contributions from the components $f_n=(u {\bar u}+d {\bar d})/\sqrt{2}$ and $f_s=s {\bar s}$ respectively.
}
\label{tab:fnfs}
\begin{ruledtabular}  \begin{threeparttable}
\setlength{\tabcolsep}{1mm}{ \begin{tabular}[t]{lcc }
Decay Modes      & $ {\cal B}_{f_n}$     &${\cal B}_{f_s}$ \\\hline
$B^0\to \rho^0 f_0\to (\pi^+\pi^-)(\pi^+\pi^-)$      &$2.29\times 10^{-8}$      & $2.37\times 10^{-9}$  \\
\end{tabular}}
\end{threeparttable}
\end{ruledtabular}
\end{table}

\begin{table}[!htbh]
\caption{Branching ratios of the considered $S$-wave four-body  $B_{(s)}\rightarrow [VS,SS]\to (\pi\pi)(\pi\pi)$ decays,
which $S=f_0(980)$ and $V=\rho$.
The world averages of experimental data are taken from PDG 2022~\cite{pdg2020}.
The sources of the theoretical errors are the same as in Table~\ref{tab:brfour}.}
\label{tab:brswave}
\begin{ruledtabular}  \begin{threeparttable}
\setlength{\tabcolsep}{1mm}{ \begin{tabular}[t]{lcc }
Decay Modes      & Branching ratio     &Data   ($10^{-7}$) \\\hline
$B^+\to \rho^+ f_0\to (\pi^+\pi^0)(\pi^+\pi^-)$      &$(1.33^{+0.54+0.20+0.13 }_{-0.35-0.17-0.14})\times 10^{-6}$      & $<20$    \\
$B^0\to \rho^0 f_0\to (\pi^+\pi^-)(\pi^+\pi^-)$      &$(0.27^{+0.12+0.12+0.07 }_{-0.08-0.03-0.04})\times 10^{-7}$      & $7.8\pm 2.5$  \\
$B^0\to f_0 f_0\to (\pi^+\pi^-)(\pi^+\pi^-)$      &$(7.42^{+5.24+3.16+3.91 }_{-2.93-0.00-3.20})\times 10^{-10}$         & $<1.9$   \\
$B_s^0\to \rho^0 f_0\to (\pi^+\pi^-)(\pi^+\pi^-)$      &$(6.80^{+3.05+1.86+0.97 }_{-2.00-1.50-0.94})\times 10^{-8}$&     \\
$B_s^0\to f_0 f_0\to (\pi^+\pi^-)(\pi^+\pi^-)$      &$(3.09^{+1.91+0.14+1.35 }_{-1.16-0.21-1.08})\times 10^{-7}$       &  $\cdots$    \\
\end{tabular}}
\end{threeparttable}
\end{ruledtabular}
\end{table}

\begin{figure}[tbp]
	\centering
	\includegraphics[scale=0.32]{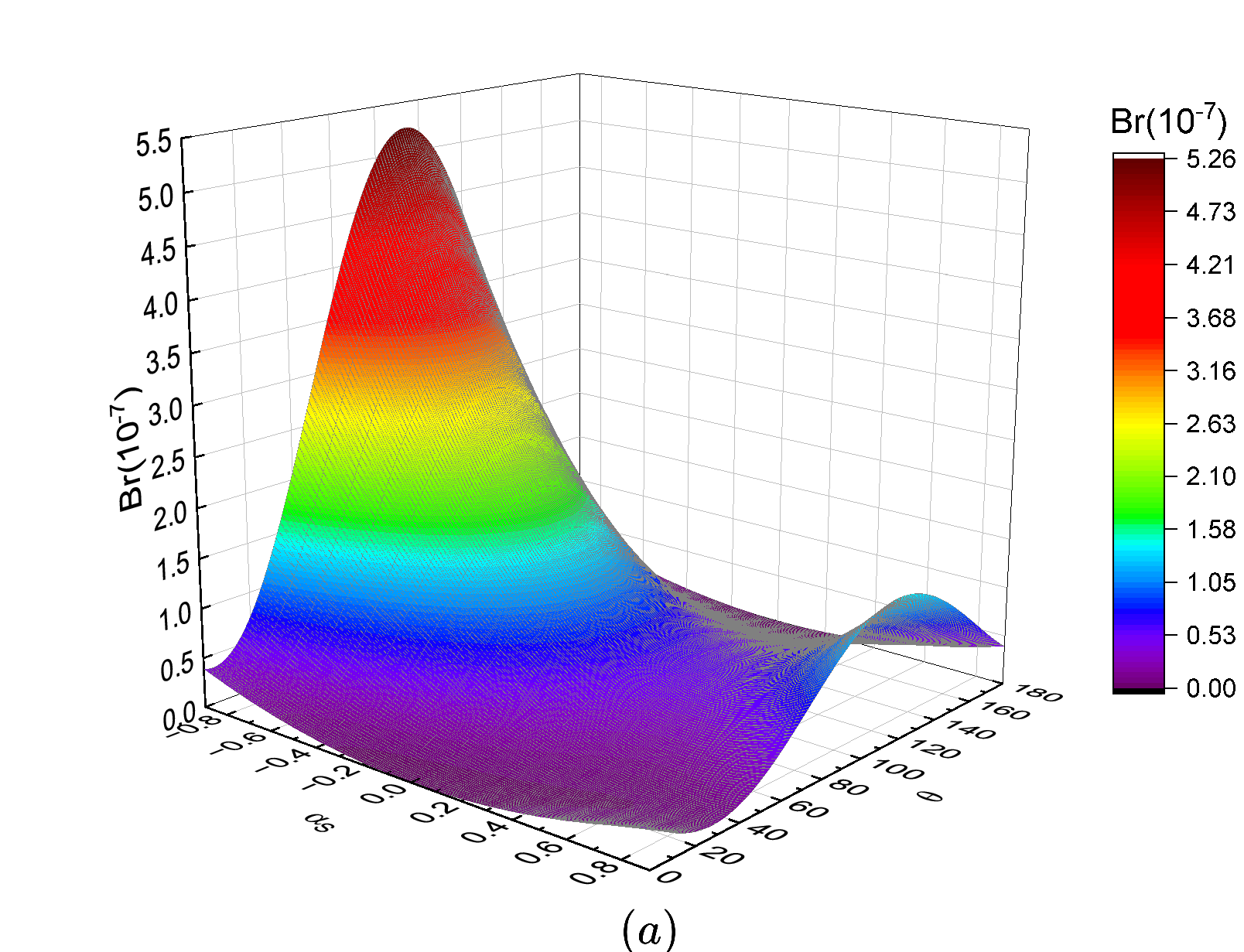}
\includegraphics[scale=0.21]{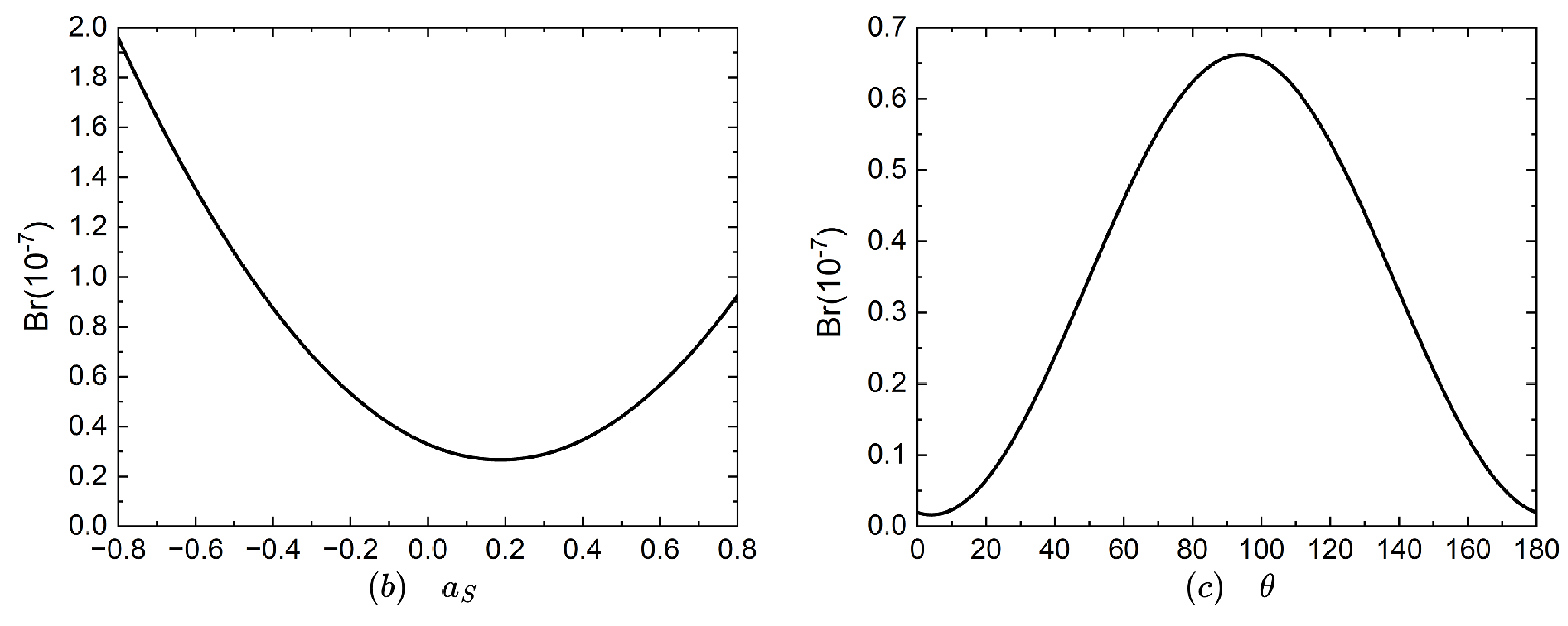}
	\caption{The branching ratio of the four-body decay $B^0\to \rho^0 f_0(980)\to (\pi^+\pi^-)(\pi^+\pi^-)$ as a function of:
$(a)$ the Gegenbauer moment $a_s$ and mixing angel $\theta$,
$(b)$ Gegenbauer moment $a_s$, and
$(c)$ mixing angel $\theta$.
The parameters $\theta=145^\circ$ and $a_S=0.2$ are fixed in the above diagrams $(b)$ and $(c)$ respectively. }
\label{fs}
\end{figure}

The branching fractions of the $S$-wave four-body decays $B_{(s)}\to [VS,SS] \to (\pi\pi)(\pi\pi)$ together with the experimental data are displayed in Table \ref{tab:brswave}.
It is worth of stressing that there already exist some well known results for $B_{(s)}\to [VS,SS]$ in the two-body framework both
in the PQCD~\cite{prd102-116007,epjc82-177,prd82-034036} and QCDF~\cite{prd77-014034,prd87-114001} approaches.
Based on the narrow width limit in Eq.~(\ref{2body}) with ${\cal B}(f_0\to \pi^+\pi^-)=0.5$ \cite{prd87-114001}, we can evaluate
${\cal B }((B^0\to \rho^0f_0\to (\pi^+\pi^-)(\pi^+\pi^-))_{\rm PQCD}=(1.65^{+1.00}_{-0.84})\times 10^{-7}$ and
${\cal B }((B^0\to \rho^0f_0\to (\pi^+\pi^-)(\pi^+\pi^-))_{\rm QCDF}=(0.10^{+0.15}_{-0.00})\times 10^{-7}$
from the previous two-body results~\cite{prd82-034036,prd87-114001}, which is close to our four-body prediction given in Table \ref{tab:brswave} within errors.
However, the known theoretical predictions are much smaller than the available data.
Maybe there are some reasons for this phenomena.
On the one hand, the tree operator contributions of the $B^0\to f_0\rho^0\to (\pi^+\pi^-)(\pi^+\pi^-)$ decay proportional to the term $(C_1+C_2/3)$ is color suppressed as shown in Eq.~(\ref{f0r0}), which leads to a small decay rate.
For such $``C"$-type decay mode, their rates are more sensitive to the NLO corrections, which is similar with those two-body decays $B^0 \to \pi^0\pi^0, \pi^0\rho^0, \rho^0\rho^0$~\cite{prd72-114005,epjc72-1923,prd73-114014,cpc46-123103}.
On the other hand, the nonperturbative input parameters from the wave functions make important sense to the branching ratios.
In Fig.~\ref{fs}, we show the dependence of ${\cal B}(B^0\to \rho^0f_0\to (\pi^+\pi^-)(\pi^+\pi^-))$ on the variation of the Gegenbauer momenta $a_S$ and the mixing angel $\theta$.
Strictly speaking, we can fit the related Gegenbauer moments and mixing angel with abundant data to match the experiment in the future.
But in fact, their rates can be accommodated in the two-quark picture for $f_0(980)$ does not mean that $\bar{q}{q}$ composition should be supported.
It is too difficult to make theoretical predictions on these decay modes based on the four-quark picture for scalar resonances.
We just assume they are constituted by two quarks at this moment.
The discrepancy between the data and the theoretical
results could be clarified with the high precision both in experimental and theoretical sides.

So far, the $B^+\to \rho^+f_0\to (\pi^+\pi^0)(\pi^+\pi^-)$, $B^0\to \rho^0f_0\to (\pi^+\pi^-)(\pi^+\pi^-)$ and $B^0\to f_0f_0\to (\pi^+\pi^-)(\pi^+\pi^-)$ decays have been measured~\cite{pdg2020}.
The PQCD predictions of ${\cal B}(B^+\to \rho^+f_0\to (\pi^+\pi^0)(\pi^+\pi^-))$ and ${\cal B}(B^0\to f_0f_0\to (\pi^+\pi^-)(\pi^+\pi^-))$ are below experimental upper limit, which is expected to be test by the future LHCb and Belle-II Collaborations.

\subsection{$CP$ asymmetry observables}

The direct $CP$ asymmetry in each component can be defined as:
\begin{eqnarray}
\quad \mathcal{A}^{\text{dir}}_h=\frac{\mathcal{\bar{B}}_h-\mathcal{B}_h}{\mathcal{\bar{B}}_h+\mathcal{B}_h}.
\end{eqnarray}
In addition,
we also focus on the TPAs originating from the interference between the $CP$-odd amplitude $A_{\perp}$ and the other two $CP$-even amplitudes $A_0$, $A_{\parallel}$
in the $B_{(s)}\to \rho\rho\to (\pi\pi)(\pi\pi)$ decays.
According to Eq.~(\ref{eq:cos}), the TPAs associated with $A_{\perp}$ for the considered four-body decays are derived from the
partially integrated differential decay rates as~\cite{prd84-096013,jhep07-166}
\begin{eqnarray} \label{eq:ATs2}
\mathcal{A}_{\text{T}}^1&=&\frac{\Gamma((2\zeta_1-1)(2\zeta_2-1)\sin\varphi>0)-\Gamma((2\zeta_1-1)(2\zeta_2-1)\sin\varphi<0)}
{\Gamma((2\zeta_1-1)(2\zeta_2-1)\sin\varphi>0)+\Gamma((2\zeta_1-1)(2\zeta_2-1)\sin\varphi<0)}\nonumber\\&=&
-\frac{2\sqrt{2}}{\pi}\int d\omega_1 d\omega_2k(\omega_1)k(\omega_2)k(\omega_1,\omega_2) \text{Im}[\frac{A_{\perp}A_0^*}{\mathcal{D}}],\\
\mathcal{A}_{\text{T}}^2&=&\frac{\Gamma(\sin2\varphi>0)-\Gamma(\sin2\varphi<0)}
{\Gamma(\sin2\varphi>0)+\Gamma(\sin2\varphi<0)}\nonumber\\&=&
-\frac{4}{\pi}\int d\omega_1 d\omega_2k(\omega_1)k(\omega_2)k(\omega_1,\omega_2) \text{Im}[\frac{A_{\perp}A_{\parallel}^*}{\mathcal{D}}],
\end{eqnarray}
with the denominator
\begin{eqnarray}
\mathcal{D}=|A_0|^2+|A_{\parallel}|^2+|A_{\perp}|^2.
\end{eqnarray}
The above TPAs take the form $\text{Im}(A_{\perp}A_{0,\parallel}^*)=|A_{\perp}||A_{0,\parallel}^*|\sin(\Delta\phi+\Delta \delta)$,
where $\Delta\phi$ and $\Delta\delta$ denote the weak and strong phase differences between the amplitudes
$A_{\perp}$ and $A_{0,\parallel}$, respectively.
It should be noted that the term $\text{Im}(A_{\perp}A_{0,\parallel}^*)$ proportional to $\sin(\Delta\phi+\Delta \delta)$
can be nonzero even in the absence of  the weak phases.
Thus,
a nonzero TPA is not quite accurate to be identified as a signal of $CP$ violation.
In order to obtain a true $CP$ violation signal,
one has to compare the TPAs in the $B$ and $\bar{B}$ meson decays.
The helicity amplitude for the $CP$-conjugated process can be inferred from  Eq.~(\ref{eq:allampli})
through the following transformations:
\begin{eqnarray}
A_0\to \bar{A}_0, \quad A_{\parallel} \to \bar{A}_{\parallel}, \quad A_{\perp} \to -\bar{A}_{\perp},
\end{eqnarray}
in which the $\bar{A}_{\lambda}$ are obtained from the $A_{\lambda}$ by changing the sign of the weak phases.
The TPAs $\bar{\mathcal{A}}_{\text{T}}^i$ associated with the $CP$-conjugated process can then be defined similarly,
but with a multiplicative minus sign.
We then have the TPAs for the $CP$-averaged decay rates
\begin{eqnarray}\label{eq:tpa}
\mathcal{A}_\text{T-true}^{1(2),\text{ave}}&\equiv& \frac{[\Gamma(T_{1(2)}>0)+\bar{\Gamma}(\bar{T}_{1(2)}>0)]-[\Gamma(T_{1(2)}<0)+\bar{\Gamma}(\bar{T}_{1(2)}<0)]}
{[\Gamma(T_{1(2)}>0)+\bar{\Gamma}(\bar{T}_{1(2)}>0)]+[\Gamma(T_{1(2)}<0)+\bar{\Gamma}(\bar{T}_{1(2)}<0)]}\non
&=&B_{1(2)}\int d\omega_1 d\omega_2k(\omega_1)k(\omega_2)k(\omega_1,\omega_2)
\text{Im}[\frac{A_{\perp}A_{0(\|)}^*-\bar{A}_{\perp}\bar{A}_{0(\|)}^*}{(\mathcal{D}+\bar{\mathcal{D}})}]\propto\sin\Delta\phi\cos\Delta\delta , \label{tpatrueave}\\
\mathcal{A}_\text{T-fake}^{1(2),\text{ave}}&\equiv& \frac{[\Gamma(T_{1(2)}>0)-\bar{\Gamma}(\bar{T}_{1(2)}>0)]-[\Gamma(T_{1(2)}<0)-\bar{\Gamma}(\bar{T}_{1(2)}<0)]}
{[\Gamma(T_{1(2)}>0)+\bar{\Gamma}(\bar{T}_{1(2)}>0)]+[\Gamma(T_{1(2)}<0)+\bar{\Gamma}(\bar{T}_{1(2)}<0)]}\non
&=&B_{1(2)}\int d\omega_1 d\omega_2k(\omega_1)k(\omega_2)k(\omega_1,\omega_2)
\text{Im}[\frac{A_{\perp}A_{0(\|)}^*-\bar{A}_{\perp}\bar{A}_{0(\|)}^*}{(\mathcal{D}+\bar{\mathcal{D}})}]\propto\cos\Delta\phi\sin\Delta\delta, \label{tpafakeave}
\end{eqnarray}
for the $CP$-conjugate decay.
The explicit expressions of the $T_{1(2)}$ and ${\bar T}_{1(2)}$ together with the factors $B_{1(2)}$ are
\begin{eqnarray}
T_1&=&(2\zeta_1-1)(2\zeta_2-1)\sin\varphi,\quad T_2=\sin2\varphi, \non
{\bar T}_1&=&(2\zeta_1-1)(2\zeta_2-1)\sin{\bar \varphi},\quad {\bar T}_2=\sin2{\bar \varphi},\non
B_1&=&-\frac{2\sqrt{2}}{\pi},\quad B_2=-\frac{4}{\pi}.
\end{eqnarray}
It is shown that the $\mathcal{A}_\text{T-true}^{1(2),\text{ave}}$ can be nonzero only in the presence of the weak phase difference,
and then provide an alternative measurement of $CP$ violation.
What's more,
compared with direct $CP$ asymmetries,
$\mathcal{A}_\text{T-true}^{1(2),\text{ave}}$ does not suffer the suppression from the strong phase difference,
and is maximal when the strong phase difference vanishes.
The $\mathcal{A}_\text{T-fake}^{1(2),\text{ave}}$, on the contrary,  can be nonzero when the weak phase difference vanishes.
Such a quantity is usually referred to as a ``fake" asymmetry ($CP$ conserving),
and simply reflects the effect of strong phases~\cite{prd84-096013,plb701-357}, instead of $CP$ violation.

For a direct comparison with the future measurements, one can also define the so-called ``true"  and ``fake"  TPAs as follows,
\begin{eqnarray}
{\cal A}_{\text{T-true}}^{1(2)}&=&\frac{1}{2}({\cal A}_{\text{T-true}}^{1(2)}+\bar{{\cal A}}_{\text{T-true}}^{1(2)})=B^{\prime}_{1(2)}\int d\omega_1 d\omega_2k(\omega_1)k(\omega_2)k(\omega_1,\omega_2) \text{Im}[\frac{A_{\perp}A_{0(||)}^*}{\mathcal{D}}-\frac{\bar{A}_{\perp}\bar{A}_{0(||)}^*}{\bar{\mathcal{D}}}],\label{tpatrue}\\
{\cal A}_\text{T-fake}^{1(2)}&=&\frac{1}{2}({\cal A}_{\text{T-fake}}^{1(2)}-\bar{{\cal A}}_{\text{T-fake}}^{1(2)})=B^{\prime}_{1(2)} \int d\omega_1 d\omega_2k(\omega_1)k(\omega_2)k(\omega_1,\omega_2) \text{Im}[\frac{A_{\perp}A_{0(||)}^*}{\mathcal{D}}+\frac{\bar{A}_{\perp}\bar{A}_{0(||)}^*}{\bar{\mathcal{D}}}] \label{tpafake},
\end{eqnarray}
with $B^{\prime}_{1}=-\frac{\sqrt{2}}{\pi}$ and $B^{\prime}_{2}=-\frac{2}{\pi}$.
Here the ``true"  and ``fake" labels refer to whether the asymmetry is due to a real $CP$ asymmetry or effects from
final-state interactions that are $CP$ symmetric.
It should be stressed that two asymmetries defined in Eqs.~(\ref{tpatrueave}) and (\ref{tpatrue}) are different in the most case, as well as two asymmetries in Eqs.~(\ref{tpafakeave}) and (\ref{tpafake}).
They become equal when no direct $CP$ violation occurs in the total rate, namely $\mathcal{D}=\bar{\mathcal{D}}$.

\begin{table}[!htbh]
\caption{
Direct $CP$ asymmetries (in units of $\%$) of the four body  $B_{(s)}\to \rho\rho \to (\pi\pi)(\pi\pi) $ decays.
For a comparison, we also list the results from the previous PQCD \cite{prd91-054033},
QCDF \cite{prd80-114026}, SCET \cite{prd96-073004}, and FAT \cite{epjc77-333}.
The world averages of experimental data are taken from PDG 2022~\cite{pdg2020}.
The sources of the theoretical errors are the same as in Table~\ref{tab:brfour}.}
\label{tab:cpvv}
\begin{ruledtabular}  \begin{threeparttable}
\setlength{\tabcolsep}{1mm}{ \begin{tabular}[t]{lcccc }
Decay Modes      & ${\cal A}^{\rm CP}_{0}$       &${\cal A}^{\rm CP}_{\|}$  & ${\cal A}^{\rm CP}_{\bot}$ & ${\cal A}^{\rm CP}$
                   \\\hline
$B^+\to\rho^+\rho^0\to(\pi^+\pi^0)(\pi^+\pi^-)$         &$0.10^{+0.01+0.04+0.03}_{-0.05-0.04-0.03}$ &$-0.62^{+0.01+0.71+0.32}_{-0.02-0.32-0.07}$ &$-0.46^{+0.06+0.51+0.16}_{-0.03-0.68-0.15}$ &$0.08^{+0.02+0.04+0.02 }_{-0.06-0.04-0.03}$\\
PQCD (former) &$0.002\pm 0.003$   &$\cdots$    &$-0.32^{+0.25}_{-0.64}$     &$0.05^{+0.06 }_{-0.03}$   \\
QCDF  &$\cdots$&$\cdots$&$\cdots$&$0.06$\\
SCET&$\cdots$&$\cdots$&$\cdots$&$0$\\
FAT &0&$\cdots$ &0 &0\\
Data &$\cdots$&$\cdots$&$\cdots$&$-5\pm 5$\\ \hline
$B^0\to\rho^+\rho^-\to(\pi^+\pi^0)(\pi^-\pi^0)$        &$-2.9^{+0.8+2.9+1.0}_{-1.1-0.8-1.5}$    &$43.4^{+5.1+8.8+6.7}_{-5.2-16.2-6.3}$
&$38.3^{+4.8+9.6+6.2}_{-4.5-15.8-5.3}$ &$-0.2^{+0.4+0.9+0.9 }_{-0.6-0.9-0.9}$\\
PQCD (former) &$-2.05^{+0.53}_{-0.55}$ &$\cdots$ &$39.0^{+7.6}_{-8.4}$ &$0.83^{+0.83 }_{-0.74}$ \\
QCDF &$\cdots$&$\cdots$&$\cdots$&$-4\pm 3$\\
SCET&$-0.07\pm 0.09$&$\cdots$&$7.68\pm 9.19$&$-7.68\pm 9.19$\\
FAT &$1.30\pm 0.54$&$\cdots$&$-16.3\pm8.2$&$-8.10\pm 2.94$\\
Data &$\cdots$&$\cdots$&$\cdots$&$0\pm 9$\\ \hline
$B^0\to\rho^0\rho^0\to(\pi^+\pi^-)(\pi^+\pi^-)$      &$85.5^{+1.0+6.7+3.8}_{-5.4-14.4-11.2}$     &$64.3^{+3.7+5.3+6.1}_{-5.8-1.1-5.8}$
 &$77.5^{+4.9+4.3+1.1}_{-5.1-5.5-1.5}$ &$77.3^{+2.2+3.8+3.2}_{-2.2-6.5-7.0}$ \\
PQCD (former) &$88.9^{+9.0}_{-120.7}$    &$\cdots$    &$-11.6^{+16.2}_{-2.9}$    &$70.7^{+4.8 }_{-9.6}$\\
QCDF &$\cdots$&$\cdots$&$\cdots$&$30^{+22 }_{-31}$\\
SCET &$2.87\pm 4.00$&$\cdots$&$-19.5\pm 23.5$&$19.5 \pm 23.5$\\
FAT  &$10.5\pm 9.6$&$\cdots$&$-46.9\pm 13.9$&$49.7\pm 13.4$\\
Data &$\cdots$&$\cdots$&$\cdots$&$20\pm 90$\\ \hline
$B_s^0\to\rho^+\rho^-\to(\pi^+\pi^0)(\pi^-\pi^0)$       &$5.2^{+0.6+0.3+1.3}_{-0.0-0.4-1.4}$      &$3.3^{+0.0+1.2+0.8}_{-1.1-2.7-1.5}$
  &$4.3^{+0.4+8.9+1.6}_{-0.0-5.2-0.9}$  &$5.2^{+0.5+1.3+0.1 }_{-0.2-1.4-0.1}$\\
PQCD (former) &$\sim 0.0$   & $\cdots$   &$30.5^{+15.0}_{-16.3}$    &$-2.9\pm 1.7$  \\
FAT &$\cdots$&$\cdots$&$\cdots$&0\\
$B_s^0\to\rho^0\rho^0\to(\pi^+\pi^-)(\pi^+\pi^-)$      &$5.2^{+0.6+0.3+1.3}_{-0.0-0.4-1.4}$      &$3.3^{+0.0+1.2+0.8}_{-1.1-2.7-1.5}$
    &$4.3^{+0.4+8.9+1.6}_{-0.0-5.2-0.9}$ &$5.2^{+0.5+1.3+0.1 }_{-0.2-1.4-0.1}$\\
PQCD (former) &$\sim 0.0$  & $\cdots$   &$30.5^{+15.0}_{-16.3}$    &$-2.9\pm 1.7$  \\
FAT &$\cdots$&$\cdots$&$\cdots$&0\\
\end{tabular}}
\end{threeparttable}
\end{ruledtabular}
\end{table}

\begin{table}[!htbh]
\caption{Direct $CP$ asymmetries (in units of $\%$) of the $S$-wave four body  $B_{(s)}\to [VS, SS] \to (\pi\pi)(\pi\pi) $ decays, with $S=f_0(980)$ and $V=\rho$.
The sources of the theoretical errors are the same as in Table~\ref{tab:brfour}.}
\label{tab:cps}
\begin{ruledtabular}  \begin{threeparttable}
\setlength{\tabcolsep}{1mm}{ \begin{tabular}[t]{lc }
Decay Modes      & PQCD  \\\hline
$B^+\to \rho^+ f_0\to (\pi^+\pi^0)(\pi^+\pi^-)$      &$-6.4^{+0.2+0.5+1.7 }_{-0.4-0.5-1.3}$         \\
$B^0\to \rho^0 f_0\to (\pi^+\pi^-)(\pi^+\pi^-)$      &$4.9^{+1.3+14.0+19.0 }_{-1.7-13.3-13.1}$         \\
$B^0\to f_0 f_0\to (\pi^+\pi^-)(\pi^+\pi^-)$      &$-52.8^{+7.2+39.3+11.5}_{-11.2-30.0-9.4}$         \\
$B_s^0\to \rho^0 f_0\to (\pi^+\pi^-)(\pi^+\pi^-)$      &$5.2^{+1.3+6.5+0.5 }_{-1.3-10.6-1.1}$         \\
$B_s^0\to f_0 f_0\to (\pi^+\pi^-)(\pi^+\pi^-)$      &$0.9^{+0.0+0.0+0.0 }_{-1.2-1.8-2.1}$         \\
\end{tabular}}
\end{threeparttable}
\end{ruledtabular}
\end{table}

The direct $CP$ asymmetries of the $B_{(s)}\to \rho\rho \to (\pi\pi)(\pi\pi)$ decays with each helicity component together with those summed over all three components are listed in Table~\ref{tab:cpvv}.
For comparison, the theoretical predictions from the previous PQCD~\cite{prd91-054033}, QCDF~\cite{prd80-114026}, SCET~\cite{prd96-073004}
and FAT~\cite{epjc77-333} in two-body framework and the experimental measurements~\cite{pdg2020} are also presented.
We show direct $CP$ asymmetries of the $S$-wave decays $B_{(s)}\to [\rho f_0,f_0f_0] \to (\pi\pi)(\pi\pi)$ in Table~\ref{tab:cps}.
It is seen that our results can accommodate the experimental data within large uncertainties.
The kinematics of the two-body decays is fixed, while the quasi-two-body decay amplitudes depend on the invariant mass of the final-state pairs and result in the differential distribution of direct $CP$ asymmetries.
It should be noticed that the finite width of the intermediate resonance appearing in the time-like form factor $F(\omega^2)$ can moderate the $CP$ asymmetry in the four-body framework.
Thus, it is reasonable to see the differences of direct $CP$ asymmetries between the two-body and four-body frameworks in the PQCD approach.

Since direct $CP$ asymmetry is proportional to the interference between the tree and penguin contributions,
a sizable interference can make the $CP$ asymmetry relatively large.
For the color allowed tree-dominant or penguin-dominant processes,
the calculated direct $CP$ asymmetries are usually expected to be small.
Taking the three measured decays $B^0 \to \rho^+\rho^-\to (\pi^+\pi^0)(\pi^-\pi^0) $, $B^+\to \rho^+\rho^0\to(\pi^+\pi^0)(\pi^+\pi^-) $
and $B^0\to \rho^0\rho^0\to (\pi^+\pi^-)(\pi^+\pi^-)$ presented in Table~\ref{tab:cpvv} as examples,
the ${\cal A}^{\rm CP}$ of the tree-dominant channels $B^0 \to \rho^+\rho^-\to (\pi^+\pi^0)(\pi^-\pi^0) $ and
$B^+\to \rho^+\rho^0\to(\pi^+\pi^0)(\pi^+\pi^-) $  are small.
Especially for the $B^+\to \rho^+\rho^0\to(\pi^+\pi^0)(\pi^+\pi^-) $ decay,
because there is no QCD-penguin contribution and the only left electroweak penguin amplitudes are rather negligible,
the predicted  ${\cal A}^{\rm CP}=(0.08^{+0.05}_{-0.07})\%$ is close to zero.
For the ``Color-suppressed"  decay $B^0 \to \rho^0\rho^0 \to (\pi^+\pi^-)(\pi^+\pi^-)$,
the large penguin contributions from the chirally enhanced annihilation diagrams are at the same level as the tree contributions from the emission diagrams,
making the  $CP$ asymmetry  as large as ${\cal A}^{\rm CP}\sim 80\%$.
To be more specific,
we show the ${\cal B}$ of the $B^+\to \rho^+\rho^0\to(\pi^+\pi^0)(\pi^+\pi^-) $ and $B^0 \to \rho^0\rho^0 \to (\pi^+\pi^-)(\pi^+\pi^-)$
from tree and penguin amplitudes respectively,
\begin{eqnarray}
{\cal B }(B^+\to \rho^+\rho^0\to(\pi^+\pi^0)(\pi^+\pi^-))=
\left\{\begin{array}{ll}
12.91\times 10^{-6}                &{\rm tree},\\
8.35\times 10^{-9}                 &{\rm penguin},\\
\end{array} \right.
\end{eqnarray}
\begin{eqnarray}
{\cal B }(B^0 \to \rho^0\rho^0 \to (\pi^+\pi^-)(\pi^+\pi^-))=
\left\{\begin{array}{ll}
1.92 \times 10^{-7}                &{\rm tree},\\
1.14 \times 10^{-7}                 &{\rm penguin}.\\
\end{array} \right.
\end{eqnarray}
It is easy to see that the contributions of the penguin diagrams in the $B^+\to \rho^+\rho^0\to(\pi^+\pi^0)(\pi^+\pi^-)$ channel
is smaller than tree ones by three orders.
While for $B^0 \to \rho^0\rho^0 \to (\pi^+\pi^-)(\pi^+\pi^-)$ decay,
the total branching ratios from the tree and penguin operators are actually close to each other.
Therefore, the direct $CP$ asymmetry of the $``C"$-type decay  $B^0 \to \rho^0\rho^0 \to (\pi^+\pi^-)(\pi^+\pi^-)$
is much larger than that of the color allowed tree-dominant decay $B^+\to \rho^+\rho^0\to(\pi^+\pi^0)(\pi^+\pi^-)$.

By comparing the numerical calculations as listed in Table~\ref{tab:cpvv},
the PQCD, QCDF, and SCET predictions for the direct $CP$ asymmetries are indeed quite different,
due to the very large difference in the mechanism to induce the $CP$ asymmetries.
As we know,
the direct $CP$ asymmetry depends on both the strong phase and the weak CKM phase.
In the SCET, only the long-distance charming penguin  is able to afford the large strong phase at leading power and leading order,
while in the QCDF and PQCD approaches, the strong phase comes from the hard spectator scattering and annihilation diagrams respectively.
Besides, the power corrections such as penguin annihilation, which are essential to resolve the $CP$ puzzles in the QCDF, are often plagued by the endpoint
divergence that in turn break the factorization theorem \cite{prd80-114008}.
In the PQCD approach, the endpoint singularity is cured by including the parton's transverse momentum.
We hope that the future experimental data with high precision can help us to differentiate three approaches.

\begin{table}[t]
\caption{PQCD predictions for the TPAs ($\%$) of the four-body $B_{(s)} \to \rho\rho \to (\pi\pi)(\pi\pi)$ decays.
The sources of theoretical errors are same as in Table~\ref{tab:brfour} but added in quadrature.}
\label{tab:tpas}
\begin{center}
\begin{threeparttable}
\begin{tabular}{l|c|c|c|c|c|c}
\hline\hline
\multicolumn{1}{c|}{}  &\multicolumn{6}{c}{ $\text{TPAs}$-1}   \cr\cline{2-7}
{Modes} &$\mathcal{A}_{\text{T}}^1$&$\bar{\mathcal{A}}_{\text{T}}^1$&$\mathcal{A}_{\text{T-true}}^1$ &$\mathcal{A}_{\text{T-fake}}^1$&$\mathcal{A}_{\text{T-True}}^{1, \text{ave}}$ &$\mathcal{A}_{\text{T-fake}}^{1, \text{ave}}$  \cr \hline
$B^+ \to \rho^+\rho^0\to (\pi^+\pi^0)(\pi^+\pi^-)$ &$-1.05^{+1.39}_{-1.38}$&$1.12^{+1.35}_{-1.36}$&$0.04\pm 0.01$
                                                   &$-1.09^{+1.39}_{-1.33}$&$0.03\pm 0.01$&$-1.08^{+1.40}_{-1.33}$\\ \hline

$B^0 \to \rho^+\rho^-\to (\pi^+\pi^0)(\pi^-\pi^0)$ &$-0.34^{+1.87}_{-2.01}$&$-0.45^{+1.24}_{-1.06}$&$-0.40^{+0.81}_{-0.80}$
                                                   &$0.06^{+1.36}_{-1.54}$&$-0.40^{+0.82}_{-0.79}$&$0.05^{+1.35}_{-1.53}$\\ \hline

$B^0 \to \rho^0\rho^0\to (\pi^+\pi^-)(\pi^+\pi^-)$ &$32.88^{+2.83}_{-2.48}$&$-17.04^{+7.58}_{-6.32}$&$7.92^{+4.63}_{-5.21}$
                                                   &$24.96^{+7.67}_{-4.65}$&$25.26^{+4.22}_{-2.37}$&$30.62^{+4.45}_{-3.07}$\\ \hline

$B_s^0 \to \rho^+\rho^-\to (\pi^+\pi^0)(\pi^-\pi^0)$ &$-0.91^{+0.88}_{-0.74}$&$0.71^{+0.73}_{-0.87}$&$-0.10^{+0.11}_{-0.13}$
                                                   &$-0.81^{+0.87}_{-0.72}$&$-0.14^{+0.12}_{-0.13}$&$-0.82^{+0.88}_{-0.72} $\\ \hline

$B_s^0 \to \rho^0\rho^0\to (\pi^+\pi^-)(\pi^+\pi^-)$ &$-0.91^{+0.88}_{-0.74}$&$0.71^{+0.73}_{-0.87}$&$-0.10^{+0.11}_{-0.13}$
                                                   &$-0.81^{+0.87}_{-0.72}$&$-0.14^{+0.12}_{-0.13}$&$-0.82^{+0.88}_{-0.72} $\\ \hline
\hline
\multicolumn{1}{c|}{}  &\multicolumn{6}{|c}{$\text{TPAs}$-2}   \cr\cline{2-7}
{Modes} &$\mathcal{A}_{\text{T}}^2$&$\bar{\mathcal{A}}_{\text{T}}^2$ &$\mathcal{A}_{\text{T-true}}^2$ &$\mathcal{A}_{\text{T-fake}}^2$&$\mathcal{A}_{\text{T-True}}^{2, \text{ave}}$ &$\mathcal{A}_{\text{T-fake}}^{2, \text{ave}}$  \cr \hline
$B^+ \to \rho^+\rho^0\to (\pi^+\pi^0)(\pi^+\pi^-)$ &$0.202^{+0.13}_{-0.14}$&$-0.209^{+0.14}_{-0.13}$&$-0.004^{+0.003}_{-0.002}$
                                                   &$0.206\pm 0.001$&$-0.003 \pm 0.002$&$0.205\pm 0.001$\\ \hline

$B^0 \to \rho^+\rho^-\to (\pi^+\pi^0)(\pi^-\pi^0)$ &$-0.34^{+0.28}_{-0.27}$&$0.25^{+0.24}_{-0.19}$&$-0.05^{+0.06}_{-0.04}$
                                                   &$-0.30^{+0.23}_{-0.25}$&$-0.04^{+0.06}_{-0.04}$&$-0.29^{+0.22}_{-0.24} $\\ \hline

$B^0 \to \rho^0\rho^0\to (\pi^+\pi^-)(\pi^+\pi^-)$ &$-7.88^{+3.59}_{-4.74}$&$19.23^{+2.49}_{-2.44}$&$5.68^{+1.55}_{-2.19}$
                                                   &$-13.55^{+2.62}_{-3.12}$&$-4.32^{+1.58}_{-2.04}$&$-9.37^{+2.66}_{-3.20}$\\ \hline

$B_s^0 \to \rho^+\rho^-\to (\pi^+\pi^0)(\pi^-\pi^0)$ &$-0.026^{+0.069}_{-0.033}$&$0.043^{+0.037}_{-0.040}$&$0.009^{+0.004}_{-0.007}$
                                                   &$-0.035^{+0.032}_{-0.033}$&$0.006^{+0.004}_{-0.005}$&$-0.033\pm 0.031 $\\ \hline

$B_s^0 \to \rho^0\rho^0\to (\pi^+\pi^-)(\pi^+\pi^-)$ &$-0.026^{+0.069}_{-0.033}$&$0.043^{+0.037}_{-0.040}$&$0.009^{+0.004}_{-0.007}$
                                                   &$-0.035^{+0.032}_{-0.033}$&$0.006^{+0.004}_{-0.005}$&$-0.033\pm 0.031 $\\ \hline

\hline\hline
\end{tabular}
\end{threeparttable}
\end{center}
\end{table}

The calculated TPAs for the four-body $B_{(s)}\to \rho\rho \to (\pi\pi)(\pi\pi)$ decays  have been  collected in Table~\ref{tab:tpas}.
As mentioned above,
the TPAs ${\cal A}^{1(2)}_{\text{T-true}}, {\cal A}^{1(2)}_{\text{T-fake}}$ are usually not equal to the average ones ${\cal A}^{1(2),\text{ave}}_{\text{T-true}}, {\cal A}^{1(2),\text{ave}}_{\text{T-fake}}$ when the decay channel has a nonzero $CP$ asymmetry.
As the total direct $CP$ asymmetry does not exceed a few percent for most of the $B_{(s)}\to \rho\rho \to (\pi\pi)(\pi\pi)$ decays,
the relations of ${\cal A}^{1(2),\text{ave}}_{\text{T-true}}\approx {\cal A}^{1(2)}_{\text{T-true}}$
and ${\cal A}^{1(2),\text{ave}}_{\text{T-fake}} \approx {\cal A}^{1(2)}_{\text{T-fake}}$
 hold in these decays.
For the color-suppressed decay $B^0\to \rho^0\rho^0 \to (\pi^+\pi^-)(\pi^+\pi^-)$,
the predicted ${\cal A}^{1(2)}_{\text{T-true}}, {\cal A}^{1(2)}_{\text{T-fake}}$ are clearly distinct
from the averaged TPAs ${\cal A}^{1(2),\text{ave}}_{\text{T-true}}, {\cal A}^{1(2),\text{ave}}_{\text{T-fake}}$
because of the large direct $CP$ violation ${\cal A}^{\rm CP}\sim 80\%$.

One can observe that the PQCD calculations of the most ``true"  TPAs
for the $B_{(s)}\to \rho\rho \to (\pi\pi)(\pi\pi)$ decays are indeed not large in the SM, less than $1\%$.
If such asymmetries with large magnitudes are observed in the future LHCb and Belle-II experiments, it is probably a signal of new physics.
However, the average ``true" TPA ${\cal A}_{\text{T-true}}^{1, \text{ave}}=(25.26^{+4.22}_{-2.37})\%$ of the color-suppressed decay $B^0\to \rho^0\rho^0\to (\pi^+\pi^-)(\pi^+\pi^-)$ is predicted to be relatively large,
and wait for the confrontation with future data.

As mentioned in the previous section,
the decay amplitude $A_{||}$ associated with transverse polarization  is always power suppressed compared with
the longitudinal component $A_0$ in the factorization assumption.
The ${\cal A}_{\text{T}}^2$ proportional to $|A_{\perp}||A^*_{\parallel}|$ is thus expected to be smaller than ${\cal A}_{\text{T}}^1$ as can be seen easily from Table \ref{tab:tpas}.
Besides,
the smallness of ${\cal A}_{\text{T}}^2$  is also attributed to the suppression from the strong phase difference between the perpendicular and parallel polarization amplitudes,
which have been supported by the previous PQCD studies~\cite{prd91-054033}.
The measurement of a large ${\cal A}_{\text{T}}^2$ would point clearly towards the presence of new physics beyond the SM.
As ``fake" TPAs are due to strong phases and require no $CP$ violation,
the large predicted ``fake" TPA ${\cal A}_{\text{T-fake}}^{1, \text{ave}}=(30.62^{+4.45}_{-3.07})\%$ of the $B^0\to \rho^0\rho^0\to (\pi^+\pi^-)(\pi^+\pi^-)$ decay
simply indicates the importance of the strong final-state phases.
We hope the future experiments can test our predictions.

\section{Conclusion}\label{sec:4}
In this work,
the four-body decays $B_{(s)} \to (\pi\pi)(\pi\pi)$ have been systematically studied under the quasi-two-body framework in the PQCD approach,
where the vector $\rho$ resonance and scalar $f_0(980)$ resonance dominate in the $\pi\pi$ invariant-mass spectrum.
The strong dynamics of the scalar or vector resonance decays into the meson pair has been parameterized into the corresponding two-meson DAs,
which has been well established in three-body $B$ meson decays.
We have improved the global fitting results of the $P$-wave longitudinal two-pion DAs
by taking the additional four-body $B^0\to \rho^+\rho^- \to (\pi^+\pi^0)(\pi^-\pi^0)$ decay into account.
With the updated two-pion DAs,
the branching ratios, polarization fractions, direct $CP$ asymmetries, as well as the TPAs
of the considered $B_{(s)} \to [\rho\rho, \rho f_0, f_0f_0] \to (\pi\pi)(\pi\pi)$ decays have been examined.

We have extracted the two-body $B_{(s)}^0 \rightarrow \rho\rho$ branching ratios from the
corresponding four-body decays under the narrow-width limit,
and shown the polarization fractions of the related decay channels.
The obtained results of the $B^0\to \rho^+\rho^-$ decay
agree well with the previous calculations performed in the two-body framework and the data within errors.
For other two decays $B^{+}\to \rho^{+}\rho^0$ and $B^{0}\to \rho^{0}\rho^0$,
the leading order PQCD prediction of ${\cal B}(B^{+}\to \rho^{+}\rho^0)$,
 ${\cal B}(B^{0}\to \rho^{0}\rho^0)$, $f_0(B^{0}\to \rho^0\rho^0)$
are below the current experimental measurements,
and could be resolved when the NLO corrections to four-body $B$ decays are available.

We have calculated the direct $CP$ asymmetries for the four-body $B_{(s)}\to (\pi\pi)(\pi\pi)$ decays.
It is shown that the direct $CP$ asymmetries could be large due to the sizable interference between the tree and penguin contributions,
while they are small for the tree-dominant or penguin-dominant processes.
The $CP$ asymmetry in the four-body framework is dependent on the invariant mass of the final-state pairs, which leads to the differences between the two-body and four-body frameworks in the PQCD approach.
In addition,
an angular analysis on four-body decays $B_{(s)}\to\rho\rho\to (\pi\pi)(\pi\pi)$ have been performed to obtain the TPAs in detail.
It is  found that most of the ``true" and ``fake" TPAs are indeed small, less than $1\%$.
For the colour suppressed decay $B^0\to \rho^0\rho^0\to (\pi^+\pi^-)(\pi^+\pi^-)$,
the large ``true" TPA ${\cal A}_{\text{T-true}}^{1, \text{ave}}=25.26\%$ is predicted for the first time in the PQCD approach,
which is clearly different from the so-called ``true" TPA $\mathcal{A}_\text{T-true}^1=7.92\%$ due to the large direct $CP$ violation.
Since ``fake" TPAs require no $CP$ violation,
the sizable fake ${\cal A}_{\text{T-fake}}^{1, \text{ave}}=30.62\%$
of the decay $B^0\to \rho^0\rho^0\to (\pi^+\pi^-)(\pi^+\pi^-)$ simply reflects the presence of significant final-state interactions.
Our predictions of the TPAs need to be further tested by the future LHCb and Belle-II measurements.

\begin{acknowledgments}
Many thanks to H.n.~Li for valuable discussions.
This work was supported by the National Natural Science Foundation of China under the No.~12005103, No.~12075086, No.~11775117, and No. 12105028.
DCY is also supported by the Natural Science Foundation of Jiangsu Province under Grant No.~BK20200980.
ZR is supported in part by the Natural Science Foundation of Hebei Province under Grant No.~A2019209449 and No.~A2021209002.

\end{acknowledgments}

\appendix
\section{Decay amplitudes}
In this Appendix we present the PQCD factorization formulas for the amplitudes of
the considered four-body hadronic $B$ meson decays:

\begin{itemize}
\item[]
$\bullet$ $ B \to \rho\rho\to(\pi\pi)(\pi\pi)$ decay modes ($h=0,\|,\perp$)
\begin{eqnarray}
\sqrt{2}A_h(B^+ \to\rho^+\rho^0 \to (\pi^+\pi^0)(\pi^+\pi^-))
&=& \frac{G_F} {\sqrt{2}}V_{ub}^*V_{ud}\Big[ \left(C_2+\frac{C_1}{3}+C_1+\frac{C_2}{3} \right )F^{LL,h}_{e\rho}+ \left(C_1+C_2\right ) M^{LL,h}_{e\rho} \Big ] \non
 &-&\frac{G_F} {\sqrt{2}}V_{tb}^*V_{td}\Big[\left(\frac{3C_9}{2}+\frac{C_{10}}{2}+\frac{3C_{10}}{2}+\frac{C_{9}}{2}
         +\frac{3C_{7}}{2}+\frac{C_{8}}{2}\right ) F^{LL,h}_{e\rho} \non
 &+& \frac{3(C_9+C_{10})}{2} M^{LL,h}_{e\rho}+\frac{3C_7}{2} M^{LR,h}_{e\rho}+\frac{3C_8}{2} M^{SP,h}_{e\rho}\Big ] ,
\end{eqnarray}
\begin{eqnarray}
A_h(B^0 \to\rho^+\rho^- \to (\pi^+\pi^0)(\pi^-\pi^0))
&=& \frac{G_F} {\sqrt{2}}V_{ub}^*V_{ud}\Big[ \left(C_2+\frac{C_1}{3} \right )F^{LL,h}_{e\rho}+ C_1 M^{LL,h}_{e\rho}\non
&+&   \left(C_1+\frac{C_2}{3} \right )F^{LL,h}_{a\rho}+ C_2 M^{LL,h}_{a\rho} \Big ] \non
 &-&\frac{G_F} {\sqrt{2}}V_{tb}^*V_{td}\Big[\left(C_4+\frac{C_3}{3}+C_{10}+\frac{C_9}{3} \right ) F^{LL,h}_{e\rho} \non
 &+& \left( C_4+C_9 \right )M^{LL,h}_{e\rho}+\left (C_5+C_7 \right ) M^{LR,h}_{e\rho}\non
 &+& \left(2C_3+\frac{2C_4}{3}+C_4+\frac{C_3}{3}+\frac{C_9}{2}+\frac{C_{10}}{6}-\frac{C_{10}}{2}-\frac{C_{9}}{6} \right )F^{LL,h}_{a\rho}\non
 &+& \left(2C_5+\frac{2C_6}{3}+\frac{C_7}{2}+\frac{C_{8}}{6}\right )F^{LR,h}_{a\rho}+ \left( C_5-\frac{C_{7}}{2} \right )M^{LR,h}_{a\rho}\non
 &+& \left(C_6+\frac{C_5}{3}-\frac{C_8}{2}-\frac{C_{7}}{6}\right )F^{SP,h}_{a\rho}+ \left( 2C_6+\frac{C_{8}}{2} \right )M^{SP,h}_{a\rho} \non
 &+& \left( C_3+2C_4-\frac{C_9}{2}+\frac{C_{10}}{2}\right )M^{LL,h}_{a\rho} \Big ] ,
\\ \non
\sqrt{2}A_h(B^0 \to\rho^0\rho^0 \to (\pi^+\pi^-)(\pi^+\pi^-))
&=& \frac{G_F} {\sqrt{2}}V_{ub}^*V_{ud}\Big[ -\left(C_1+\frac{C_2}{3} \right )F^{LL,h}_{e\rho}- C_2 M^{LL,h}_{e\rho}\non
&+&   \left(C_1+\frac{C_2}{3} \right )F^{LL,h}_{a\rho}+ C_2 M^{LL,h}_{a\rho} \Big ] \non
 &-&\frac{G_F} {\sqrt{2}}V_{tb}^*V_{td}\Big[\left(C_4+\frac{C_3}{3}-\frac{3C_9}{2}-\frac{C_{10}}{2}-\frac{C_{10}}{2}-\frac{C_9}{6}-\frac{3C_7}{2}-\frac{C_{8}}{2}\right ) F^{LL,h}_{e\rho} \non
 &+& \left( C_3-\frac{C_9}{2}- \frac{3C_{10}}{2}\right )M^{LL,h}_{e\rho}-\frac{3C_{8}}{2} M^{SP,h}_{e\rho}\non
 &+& \left(2C_3+\frac{2C_4}{3}+C_4+\frac{C_3}{3}+\frac{C_9}{2}+\frac{C_{10}}{6}-\frac{C_{10}}{2}-\frac{C_{9}}{6} \right )F^{LL,h}_{a\rho}\non
 &+& \left(2C_5+\frac{2C_6}{3}+\frac{C_7}{2}+\frac{C_{8}}{6}\right )F^{LR,h}_{a\rho}+ \left( C_5-\frac{C_{7}}{2} \right )M^{LR,h}_{a\rho}\non
 &+& \left(C_6+\frac{C_5}{3}-\frac{C_8}{2}-\frac{C_{7}}{6}\right )F^{SP,h}_{a\rho}+ \left( 2C_6+\frac{C_{8}}{2} \right )M^{SP,h}_{a\rho} \non
 &+& \left( C_3+2C_4-\frac{C_9}{2}+\frac{C_{10}}{2}\right )M^{LL,h}_{a\rho} \Big ] ,
\\\non
A_h(B_s^0 \to\rho^+\rho^- \to (\pi^+\pi^0)(\pi^-\pi^0))
&=& \frac{G_F} {\sqrt{2}}V_{ub}^*V_{us}\Big[ \left(C_1+\frac{C_2}{3}\right )F^{LL,h}_{a\rho}+ C_2 M^{LL,h}_{a\rho} \Big ] \non
 &-&\frac{G_F} {\sqrt{2}}V_{tb}^*V_{ts}\Big[\left(2C_3+\frac{2C_{4}}{3}+\frac{C_{9}}{2}+\frac{C_{10}}{6}\right ) F^{LL,h}_{a\rho} \non
 &+&\left(2C_5+\frac{2C_{6}}{3}+\frac{C_{7}}{2}+\frac{C_{8}}{6}\right ) F^{LR,h}_{a\rho} +\left(2C_4+\frac{C_{10}}{2}\right ) M^{LL,h}_{a\rho}\non
 &+&\left(2C_6+\frac{C_{8}}{2}\right ) M^{SP,h}_{a\rho}\Big ] ,
\end{eqnarray}

\begin{eqnarray}
\sqrt{2}A_h(B_s^0 \to\rho^0\rho^0 \to (\pi^+\pi^-)(\pi^+\pi^-))
&=& \frac{G_F} {\sqrt{2}}V_{ub}^*V_{us}\Big[ \left(C_1+\frac{C_2}{3}\right )F^{LL,h}_{a\rho}+ C_2 M^{LL,h}_{a\rho} \Big ] \non
 &-&\frac{G_F} {\sqrt{2}}V_{tb}^*V_{ts}\Big[\left(2C_3+\frac{2C_{4}}{3}+\frac{C_{9}}{2}+\frac{C_{10}}{6}\right ) F^{LL,h}_{a\rho} \non
 &+&\left(2C_5+\frac{2C_{6}}{3}+\frac{C_{7}}{2}+\frac{C_{8}}{6}\right ) F^{LR,h}_{a\rho} +\left(2C_4+\frac{C_{10}}{2}\right ) M^{LL,h}_{a\rho}\non
 &+&\left(2C_6+\frac{C_{8}}{2}\right ) M^{SP,h}_{a\rho}\Big ].
\end{eqnarray}

According to Eq.~(\ref{f0mix}),
the total decay amplitudes of the $S$-wave channels can be divided  into the
$n {\bar n}=\frac{1}{\sqrt{2}}(u{\bar u}+d{\bar d})$ and $s{\bar s}$ components.

\item[]
$\bullet$ $ B \to f_0(980) \rho\to(\pi\pi)(\pi\pi)$ decay modes
\begin{eqnarray}
2A(B_s^0 \to f_n\rho^0\to(\pi^+\pi^-)(\pi^+\pi^-))&=&\frac{G_F} {\sqrt{2}}V_{ub}^*V_{us}\Big[\left (C_1+\frac{C_2}{3}\right)(F^{LL}_{af_n}+F^{LL}_{a\rho})
+C_2(M^{LL}_{af_n}+M^{LL}_{a\rho})\Big]\non
&-& \frac{G_F} {\sqrt{2}}V_{tb}^*V_{ts}\Big[\left (\frac{3C_{7}}{2}+\frac{C_8}{2}+\frac{3C_{9}}{2}+\frac{C_{10}}{2}\right)(F^{LL}_{af_n}+F^{LL}_{a\rho})\non
&+&\frac{3C_{10}}{2}(M^{LL}_{af_0}+M^{LL}_{a\rho})+\frac{3C_{8}}{2}(M^{SP}_{af_0}+M^{SP}_{a\rho})\Big],
\\\non
\sqrt{2}A(B_s^0 \to f_s\rho^0\to(\pi^+\pi^-)(\pi^+\pi^-))&=&\frac{G_F} {\sqrt{2}}V_{ub}^*V_{us}\Big[\left (C_1+\frac{C_2}{3}\right)F^{LL}_{ef_s}+C_2M^{LL}_{ef_s}\Big]\non
&-& \frac{G_F} {\sqrt{2}}V_{tb}^*V_{ts}\Big[\left (\frac{3C_{7}}{2}+\frac{C_8}{2}+\frac{3C_{9}}{2}+\frac{C_{10}}{2}\right)F^{LL}_{ef_s}+\frac{3C_{10}}{2}M^{LL}_{ef_s}+\frac{3C_{8}}{2}M^{SP}_{ef_s}\Big].\non
\\\non
\sqrt{2}A(B^+ \to f_n\rho^+\to(\pi^+\pi^-)(\pi^+\pi^0))&=& \frac{G_F} {\sqrt{2}}V_{ub}^*V_{ud}\Big[\left (C_2+\frac{C_1}{3}\right)(F^{LL}_{ef_n}+F^{LL}_{af_n}+F^{LL}_{a\rho})\non
&+&C_1\left ( M^{LL}_{ef_n}+M^{LL}_{af_n}+M^{LL}_{a\rho}\right )+C_2M^{LL}_{e\rho}\Big]\non
&-& \frac{G_F} {\sqrt{2}}V_{tb}^*V_{td}\Big[\left (C_4+\frac{C_3}{3}+C_{10}+\frac{C_9}{3}\right)(F^{LL}_{ef_n}+F^{LL}_{af_n}+F^{LL}_{a\rho})\non
&+& \left(C_6+\frac{C_5}{3}+C_8+\frac{C_7}{3} \right)(F^{SP}_{af_n}+F^{SP}_{a\rho})+\left( 2C_6+\frac{C_{8}}{2}\right)M^{SP}_{e\rho} \non
&+&\left(C_6+\frac{C_5}{3}-\frac{C_8}{2}-\frac{C_{7}}{6} \right)F^{SP}_{e\rho}+\left( C_5-\frac{C_7}{2}\right )M^{LR}_{e\rho}\non
&+&\left( C_3+C_9\right )(M^{LL}_{ef_n}+M^{LL}_{af_n}+M^{LL}_{a\rho})\non
&+& \left( C_5+C_7\right )(M^{LR}_{ef_n}+M^{LR}_{af_n}+M^{LR}_{a\rho})\non
&+&\left(C_3+2C_4-\frac{C_9}{2}+\frac{C_{10}}{2} \right )M^{LL}_{e\rho}\Big],\label{f0rp}
\\\non
A(B^+ \to f_s\rho^+\to(\pi^+\pi^0)(\pi^+\pi^-))&=&
- \frac{G_F} {\sqrt{2}}V_{tb}^*V_{td}\Big[\left (C_4-\frac{C_{10}}{2}\right)M^{LL}_{e\rho}+\left (C_6-\frac{C_{8}}{2}\right)M^{SP}_{e\rho}\Big],
\end{eqnarray}
\begin{eqnarray}
2A(B^0 \to f_n\rho^0\to(\pi^+\pi^-)(\pi^+\pi^-))&=& \frac{G_F} {\sqrt{2}}V_{ub}^*V_{ud}\Big[\left (C_1+\frac{C_2}{3}\right)(F^{LL}_{ef_n}+F^{LL}_{af_n}+F^{LL}_{a\rho})\non
&+&C_2\left ( M^{LL}_{ef_n}+M^{LL}_{af_n}-M^{LL}_{e\rho}+M^{LL}_{a\rho}\right )\Big]\non
&-& \frac{G_F} {\sqrt{2}}V_{tb}^*V_{td}\Big[\left (-C_4-\frac{C_3}{3}+\frac{3C_7}{2}+\frac{C_{8}}{2}+\frac{3C_9}{2}
+\frac{C_{10}}{2}+\frac{C_{10}}{2}+\frac{C_9}{6}\right)\non
&\times& \left(F^{LL}_{ef_n}+F^{LL}_{af_n}+F^{LL}_{a\rho}\right) +\left(-C_6-\frac{C_5}{3}+\frac{C_8}{2}+\frac{C_{7}}{6} \right)(F^{SP}_{e\rho}+F^{SP}_{af_n}+F^{SP}_{a\rho})\non
&-&\left( C_3-\frac{C_9}{2}-\frac{3C_{10}}{2}\right )(M^{LL}_{ef_n}+M^{LL}_{af_n}+M^{LL}_{a\rho})-\left( 2C_6+\frac{C_{8}}{2}\right)M^{SP}_{e\rho} \non
&-& \left(C_3+2C_4-\frac{C_9}{2}+\frac{C_{10}}{2} \right )M^{LL}_{e\rho}+\frac{3C_{8}}{2} (M^{SP}_{ef_n}+M^{SP}_{af_n}+M^{SP}_{a\rho})\non
&-&\left( C_5-\frac{C_7}{2}\right )(M^{LR}_{ef_n}+M^{LR}_{af_n}+M^{LR}_{e\rho}+M^{LR}_{a\rho})\Big],\label{f0r0}
\\\non
\sqrt{2}A(B^0 \to f_s\rho^0\to(\pi^+\pi^-)(\pi^+\pi^-))&=&
- \frac{G_F} {\sqrt{2}}V_{tb}^*V_{td}\Big[\left (-C_4+\frac{C_{10}}{2}\right)M^{LL}_{e\rho}+\left (-C_6+\frac{C_{8}}{2}\right)M^{SP}_{e\rho}\Big].
\end{eqnarray}

The decay amplitudes for the physical states are then
\begin{eqnarray}
A(B_{(s)} \to f_0 \rho\to(\pi\pi)(\pi\pi))=A(B_{(s)} \to f_n \rho\to(\pi\pi)(\pi\pi))\sin \theta+A(B_{(s)} \to f_s \rho\to(\pi\pi)(\pi\pi))\cos \theta.\non
\end{eqnarray}

\item[]
$\bullet$ $B \to f_0(980) f_0(980)\to(\pi\pi)(\pi\pi)$ decay modes
\begin{eqnarray}
\sqrt{2} A(B^0 \to f_n f_s\to(\pi^+\pi^-)(\pi^+\pi^-))&=& -\frac{G_F} {\sqrt{2}}V_{tb}^*V_{td}\Big[ \left(C_4-\frac{C_{10}}{2} \right )M^{LL}_{ef_n}+\left( C_6-\frac{C_{8}}{2}\right)M^{SP}_{ef_n}\Big],
\\\non
A(B^0 \to f_s f_s\to(\pi^+\pi^-)(\pi^+\pi^-))&=& -\frac{2G_F} {\sqrt{2}}V_{tb}^*V_{td}\Big[ \Big(C_3+\frac{C_4}{3}+C_5+\frac{C_6}{3} -\frac{C_7}{2}-\frac{C_8}{6}
-\frac{C_{9}}{2}-\frac{C_{10}}{6}\Big)F^{LL}_{af_s}\non
&+&\left(C_4-\frac{C_{10}}{2} \right )M^{LL}_{af_s}+\left( C_6-\frac{C_{8}}{2}\right)M^{SP}_{af_s}\Big],
\\\non
A(B^0 \to f_n f_n\to(\pi^+\pi^-)(\pi^+\pi^-))&=& \frac{G_F} {\sqrt{2}}V_{ub}^*V_{ud}\Big[\left (C_1+\frac{C_2}{3}\right)(F^{LL}_{af_n})
+C_2\left ( M^{LL}_{ef_n}+M^{LL}_{af_n}\right )\Big]\non
&-& \frac{G_F} {\sqrt{2}}V_{tb}^*V_{td}\Big[ \Big(2C_3+\frac{2C_4}{3}+C_4+\frac{C_3}{3}+2C_5+\frac{2C_6}{3}+\frac{C_7}{2}+\frac{C_8}{6}\non
&+&\frac{C_{9}}{2}+\frac{C_{10}}{6}-\frac{C_{10}}{2}-\frac{C_{9}}{6} \Big)F^{LL}_{af_n}+\left(C_6+\frac{C_5}{3}-\frac{C_8}{2}-\frac{C_{7}}{6} \right)\left(F^{SP}_{ef_n}+F^{SP}_{af_n}\right)\non
&+&\left(C_3+2C_4-\frac{C_9}{2}+\frac{C_{10}}{2} \right )\left(M^{LL}_{ef_n}+M^{LL}_{af_n}\right)+\left( C_5-\frac{C_7}{2}\right )\left(M^{LR}_{ef_n}+M^{LR}_{af_n}\right)\non
&+& \left( 2C_6+\frac{C_{8}}{2}\right)\left(M^{SP}_{ef_n}+M^{SP}_{af_n}\right) \Big],
\\\non
A(B_s^0 \to f_n f_n\to(\pi^+\pi^-)(\pi^+\pi^-))&=& \frac{G_F} {\sqrt{2}}V_{ub}^*V_{us}\Big[\left (C_1+\frac{C_2}{3}\right)(F^{LL}_{af_n})
+C_2M^{LL}_{af_n}\Big]\non
&-& \frac{G_F} {\sqrt{2}}V_{tb}^*V_{ts}\Big[ \left(2C_3+\frac{2C_4}{3}+2C_5+\frac{2C_6}{3}+\frac{C_7}{2}+\frac{C_8}{6}+\frac{C_{9}}{2}+\frac{C_{10}}{6}\right)F^{LL}_{af_n}\non
&+&\left(2C_4+\frac{C_{10}}{2} \right )M^{LL}_{af_n}+\left( 2C_6+\frac{C_{8}}{2}\right)M^{SP}_{af_n}\Big],
\end{eqnarray}
\begin{eqnarray}
\sqrt{2}A(B_s^0 \to f_n f_s\to(\pi^+\pi^-)(\pi^+\pi^-))&=& \frac{G_F} {\sqrt{2}}V_{ub}^*V_{us}\Big[C_2M^{LL}_{ef_s}\Big]\non
&-& \frac{G_F} {\sqrt{2}}V_{tb}^*V_{ts}\Big[ \left(2C_4+\frac{C_{10}}{2} \right )M^{LL}_{ef_s}+\left( 2C_6+\frac{C_{8}}{2}\right)M^{SP}_{ef_s}\Big],
\\\non
A(B^0 \to f_s f_s\to(\pi^+\pi^-)(\pi^+\pi^-))&=& -\frac{2G_F} {\sqrt{2}}V_{tb}^*V_{ts}\Big[ \Big(C_3+\frac{C_4}{3}+C_4+\frac{C_3}{3}-\frac{C_{9}}{2}-\frac{C_{10}}{6}-\frac{C_{10}}{2}-\frac{C_{9}}{6} \Big)F^{LL}_{af_s}\non
&+&\left(C_6+\frac{C_5}{3}-\frac{C_8}{2}-\frac{C_{7}}{6} \right)\left(F^{SP}_{ef_s}+F^{SP}_{af_s}\right)\non
&+&\left(C_3+C_4-\frac{C_9}{2}\frac{C_{10}}{2} \right )\left(M^{LL}_{ef_s}+M^{LL}_{af_s}\right)+\left( C_5-\frac{C_7}{2}\right )\left(M^{LR}_{ef_s}+M^{LR}_{af_s}\right)\non
&+& \left( C_6-\frac{C_{8}}{2}\right)\left(M^{SP}_{ef_s}+M^{SP}_{af_s}\right) \Big].
\end{eqnarray}

The decay amplitudes for the physical states can then be written as,
\begin{eqnarray}
\sqrt{2}A(B_{(s)} \to f_0 f_0\to(\pi\pi)(\pi\pi))&=&A(B_{(s)} \to f_n f_n\to(\pi\pi)(\pi\pi))(\sin \theta)^2\non
&+&A(B_{(s)} \to f_n f_s\to(\pi\pi)(\pi\pi))\sin 2\theta\non
&+&A(B_{(s)} \to f_s f_s\to(\pi\pi)(\pi\pi))(\cos \theta)^2.
\end{eqnarray}

\end{itemize}
$G_F=1.16639\times 10^{-5}$ GeV$^{-2}$ is the Fermi coupling constant and the $V_{ij}$'s are the Cabibbo-Kobayashi-Maskawa matrix elements.
The superscripts $LL$, $LR$, and $SP$ refer to the contributions from $(V-A)\otimes(V-A)$, $(V-A)\otimes(V+A)$, and $(S-P)\otimes(S+P)$ operators, respectively.
The explicit formulas for the factorizable emission (annihilation) contributions $F_{e(a)}$ and the nonfactorizable emission (annihilation) contributions
$M_{e(a)}$ from Fig.~\ref{fig2} can be obtained easily in Ref.~\cite{zjhep}.


\end{document}